\documentclass[reprint,superscriptaddress,amsmath,amssymb,aps,prx]{revtex4-2}

\usepackage{graphicx}
\usepackage{dcolumn}
\usepackage{bm}
\usepackage{siunitx}
\usepackage{xcolor}
\definecolor{light-gray}{gray}{0.8}
\definecolor{prlblue}{rgb}{0.176, 0.152, 0.57}
\usepackage[colorlinks=true, linkcolor=black, citecolor=prlblue, filecolor=black, urlcolor=prlblue]{hyperref}
\usepackage{multirow}

\begin{document}

\title{Achromatic optics using nonlinear plasma lenses\\ for beam-quality preservation between plasma-accelerator stages}

\author{C.~A.~Lindstr{\o}m}
\email{c.a.lindstrom@fys.uio.no}
\affiliation{Department of Physics, University of Oslo, Oslo, Norway}
\author{E.~Adli}
\affiliation{Department of Physics, University of Oslo, Oslo, Norway}
\author{J.~B.~B.~Chen}
\affiliation{Department of Physics, University of Oslo, Oslo, Norway}
\author{P.~Drobniak}
\affiliation{Department of Physics, University of Oslo, Oslo, Norway}
\author{A.~Huebl}
\affiliation{Lawrence Berkeley National Laboratory, Berkeley (CA), USA}
\author{D.~Kalvik}
\affiliation{Department of Physics, University of Oslo, Oslo, Norway}
\author{C.~E.~Mitchell}
\affiliation{Lawrence Berkeley National Laboratory, Berkeley (CA), USA}
\author{F.~Pe\~na}
\affiliation{Department of Physics, University of Oslo, Oslo, Norway}
\affiliation{Ludwig-Maximilians-Universit{\"a}t M{\"u}nchen, Munich, Germany}
\author{K.~N.~Sjobak}
\affiliation{Department of Physics, University of Oslo, Oslo, Norway}


\begin{abstract}
    Plasma acceleration promises to deliver high-energy particle beams by combining, or staging, several low- or medium-energy accelerator stages. However, chromatic aberrations from the combination of high divergence and energy spread make it nontrivial to transport beams between plasma-accelerator stages. This paper describes a compact and achromatic lattice optimized for staging, based on a new beam-optics element; a nonlinear plasma lens. The lattice preserves emittance for energy spreads up to several percent and has a tunable $R_{56}$ that enables bunch-length preservation or a longitudinal self-correction mechanism. The performance and limitations of the plasma-lens-based solution are modeled analytically and numerically, and compared to a more conventional yet novel solution based on quadrupole and sextupole magnets. While functional, the latter is double the length, has about twice the number of elements and a narrower energy bandwidth. Lastly, a solution for scaling to TeV energies is described, in which all lengths scale with the square root of the energy and the deleterious effects of coherent and incoherent synchrotron radiation are mitigated.
\end{abstract}

\maketitle



\section{Introduction}

Plasma accelerators~\cite{Esarey2009,LindstromCorde2025} can sustain orders of magnitude higher accelerating fields (1--\SI{100}{GV/m})~\cite{Hogan2005,Leemans2006} compared to radio-frequency accelerators (10--\SI{100}{MV/m})~\cite{Grudiev2009}, promising to shrink the size and cost of future particle-accelerator facilities~\cite{Joshi2003}. They operate by utilizing a charge-density wave driven by an intense laser pulse~\cite{Tajima1979} or a charged particle beam~\cite{Veksler1956,Fainberg1956,Chen1985,Ruth1985} propagating in a plasma. The resulting plasma-cavity structure has both strong accelerating fields as well as strong focusing fields, simultaneously accelerating and guiding a trailing electron bunch~\cite{Hogan2003,Muggli2004}. However, the strong focusing also results in highly diverging beams when exiting the plasma accelerator into a vacuum. This can be problematic when use of multiple plasma-accelerator stages, also known as \textit{staging}~\cite{Steinke2016,Lindstrom2021a}, is required.

Staging allows acceleration to energies higher than those achievable in a single stage. For laser- and electron-driven plasma accelerators, the energy gain is typically limited to \SI{\sim10}{GeV} per stage~\cite{Litos2016,Gonsalves2019,Picksley2024}. Higher gain may be possible in a single stage with the use of proton drivers~\cite{AdliMuggli2016,AWAKE2018,Farmer2024}, but due to challenges related to repetition rate~\cite{Darcy2022} and energy efficiency~\cite{Litos2014,Lindstrom2021c}, staging may be a necessity for high-energy, high-power applications such as a plasma-based collider~\cite{Rosenzweig1998,Seryi2009,Adli2013,Foster2023,Foster2025}. Multistage plasma acceleration could also be the key to unlocking affordable high-energy experiments in strong-field quantum electrodynamics~\cite{Gonoskov2022,Lindstrom2025}. Furthermore, use of multiple stages enables self-stabilization mechanisms in the longitudinal phase space~\cite{Lindstrom2021b}, which may be important for loosening the tight tolerances in plasma accelerators---even in cases where a single stage would in principle be sufficient.

The key challenge with staging is \textit{chromaticity}~\cite{Antici2012,Migliorati2013}, i.e.~energy-dependent focusing: since the focal length of beam optics depends on the energy of the particles being focused, refocusing a diverging beam with finite energy spread into the subsequent stage will result in the lower energy particles being focused further upstream and the higher energies further downstream. This chromaticity can be quantified in a given transverse plane in terms of a vector with an amplitude and a phase. The so-called \textit{chromatic amplitude} is given by~\cite{Montague1979}
\begin{equation}
    \label{eq:chromatic-amplitude-definition}
    W = \sqrt{\left(\frac{\partial\alpha}{\partial\delta}-\frac{\alpha}{\beta}\frac{\partial\beta}{\partial\delta}\right)^2 + \left(\frac{1}{\beta}\frac{\partial\beta}{\partial\delta}\right)^2},
\end{equation}
where $\delta$ is a relative energy offset and $\beta$ and $\alpha$ are Twiss parameters, and the \textit{chromatic phase} increases at twice the rate of the normal phase advance in the lattice. If the chromaticity is not mitigated, the corresponding relative transverse emittance growth is given by~\cite{Lindstrom2016a}
\begin{equation}
    \label{eq:chromaticity}
    \frac{\Delta\varepsilon ^2}{\varepsilon ^2} = W^2 \sigma_{\delta}^2 + \mathcal{O}(\sigma_{\delta}^4) \approx \frac{4L^2}{\beta_0^2}\sigma_{\delta}^2,
\end{equation}
where $\varepsilon $ denotes the emittance, $\sigma_{\delta}$ is the root-mean-square (rms) relative energy spread, $L$ is the distance to the refocusing optic and $\beta_0$ is the matched Twiss beta function at the exit of the plasma accelerator. This aberration is worse with higher divergence (i.e., smaller $\beta_0$) and higher energy spread. Moreover, the minimum distance $L$ is set by the space needed to separate the accelerated beam spatially from the energy-depleted driver~\cite{Pena2024} and/or merge it with a fresh driver: for laser drivers this can be done compactly with plasma mirrors~\cite{Thaury2007,Steinke2016,Zingale2021}, but for beam drivers this requires dipole dispersion (i.e., separation by energy) as no kickers are sufficiently fast.

A setup with similar challenges is that of final focusing for colliders, for which powerful chromaticity correction techniques have been developed. In principle, it is possible to cancel chromaticity with only linear optics (e.g., quadrupole magnets)~\cite{Montague1987,Lindstrom2016a}, known as apochromatic focusing, but this technique has a limited energy bandwidth. To support a larger energy spread, nonlinear optics (e.g., sextupoles) are required. An elegant such solution is \textit{local chromaticity correction}~\cite{Raimondi2001}, whereby beams are energetically dispersed with a dipole magnet into the final quadrupole doublet, each quadrupole having a sextupole magnet placed just upstream of it. Near the transverse horizontal ($x$) axis, the sextupole field can be approximated as a quadrupole field that linearly varies in strength along $x$ (e.g., stronger on the right than on the left). By matching the sextupole strength to the dispersion at that point, the combined focal length from the quadrupole and sextupole can be made equal for all energy slices, making the quadrupole--sextupole pair effectively achromatic. A second doublet of nearly identical quadrupole--sextupole pairs is placed upstream, at \SI{180}{\degree} phase advance, to cancel the unwanted off-axis nonlinear kicks induced by the sextupoles. While this technique has proven highly successful~\cite{White2014}, the setup can be long and complex due to the quantity and length of magnets required. For a multistage plasma accelerator, where the goal is to maximize the average accelerating gradient, placing long optics between stages is therefore not ideal.

Plasma lenses enable more compact focusing, due to their increased focusing gradient and because they focus in both transverse planes simultaneously, as opposed to the focusing--defocusing in $x$--$y$ seen in quadrupole magnets. Two variations exist: \textit{passive plasma lenses} (PPLs)~\cite{Nakanishi1991,Ng2001,Marocchino2017,Doss2019}, which are effectively just short plasma accelerators where only the transverse electric field is utilized; and \textit{active plasma lenses} (APLs)~\cite{Panofsky1950,vanTilborg2015,Pompili2018,Lindstrom2018a}, which make use of a longitudinal discharge current to focus the beam with an azimuthal magnetic field. Active plasma lenses can reach magnetic field gradients of order kT/m~\cite{vanTilborg2015,Sjobak2021}, about 10--100 times higher than conventional quadrupole magnets, whereas passive plasma lenses can reach field gradients as high as \SI{e15}{V/m^2}, equivalent to MT/m. However, while the stronger focusing fields allow shorter focal lengths, reducing $L$ in Eq.~\ref{eq:chromaticity} and therefore the chromaticity~\cite{vanTilborg2018}, plasma lenses are still not inherently achromatic.

In this paper, we combine the achromaticity of local chromaticity correction~\cite{Raimondi2001} and the compactness of plasma lenses into a new achromatic lattice for staging of plasma accelerators. This requires the introduction of \textit{nonlinear plasma lenses}, which have a transverse gradient in their focusing strength (e.g., stronger on the right than on the left)~\cite{Doss2023,Drobniak2025a,Drobniak2025b}, in combination with magnetic dipoles. The achromatic lattice primarily uses these dipoles for chromaticity correction and in- and out-couple the plasma-stage drivers, but also to provide a longitudinal dispersion ($R_{56}$) which facilitates multistage self-correction in the longitudinal phase space~\cite{Lindstrom2021b}---a concept similar to that of synchrotron oscillations, but which also damps energy spread and energy offsets.

The following sections describe: the required nonlinear plasma-lens field profile (Sec.~\ref{sec:nonlinear-plasma-lenses}); the optics and capabilities of the achromatic lattice (Sec.~\ref{sec:achromatic-lattice}); an alternative quadrupole-based lattice for comparison purposes (Sec.~\ref{sec:achromatic-quad-lattice}); the overall performance and limitations of these lattices (Sec.~\ref{sec:performance-and-limitations}); solutions for energy scaling and the effect of synchrotron radiation (Sec.~\ref{sec:energy-scaling-and-sr}); and finally some concluding remarks (Sec.~\ref{sec:conclusions}).


\section{Nonlinear plasma lenses}
\label{sec:nonlinear-plasma-lenses}

In order to replace both the quadrupole and sextupole typically used in local chromaticity correction, the nonlinear plasma lens must have both a quadrupole-like and a sextupole-like field component. Whereas both quadrupole and sextupole magnetic fields are curl-free, $\nabla \times \mathbf{B} = 0$, the equivalent electric or magnetic field components in a plasma lens are not. Below we derive the plasma-lens field required to be achromatic for a beam dispersed in the lens with a horizontal dispersion $D_x = \partial x/\partial \delta$, where $\delta$ is the relative energy offset of a particle compared to the nominal energy $\mathcal{E}_0$. Here we will assume that all particles are ultrarelativistic, since staging mainly applies to high-energy electrons and positrons.

Particle trajectories evolve according to the Lorentz force equation, $\mathbf{F} = q(\mathbf{E} + \mathbf{v} \times \mathbf{B})$, where $q$ and $\mathbf{v}$ are the particle charge and velocity, respectively, and $\mathbf{E}$ and $\mathbf{B}$ are the electric and magnetic fields. In a linear plasma lens, the transverse position and angle of a particle with energy $\mathcal{E}=\mathcal{E}_0(1+\delta)$ evolves according to 
\begin{eqnarray}
    \label{eq:trajectory-evol-x}
    x'' &=& -kx = \frac{F_x}{\mathcal{E}} \\
    \label{eq:trajectory-evol-y}
    y'' &=& -ky = \frac{F_y}{\mathcal{E}}, 
\end{eqnarray}
where the prime denotes the longitudinal derivative along the trajectory and $k$ is the focusing strength of the lens. While the force itself is energy-\textit{independent}, the kick (change in angle; Eqs.~\ref{eq:trajectory-evol-x} and \ref{eq:trajectory-evol-y}) and hence the focusing strength are energy-\textit{dependent}: $k=k_0/(1+\delta)$, where $k_0$ is the focusing strength at nominal energy. In order to cancel this energy dependence, we can introduce a position-dependent focusing gradient (i.e., the transverse gradient of the transverse force) that varies in $x$ as
\begin{align}
    \label{eq:force-gradient-x}
    \frac{\partial F_x}{\partial x} &= q \left(\frac{\partial E_x}{\partial x} - c \frac{\partial B_y}{\partial x}\right) = -\mathcal{E} \frac{k_0}{1+\delta} \left(1 + \tau_x x\right), \\
    \label{eq:force-gradient-y}
    \frac{\partial F_y}{\partial y} &= q \left(\frac{\partial E_y}{\partial y} + c \frac{\partial B_x}{\partial y}\right) = -\mathcal{E} \frac{k_0}{1+\delta} \left(1 + \tau_x x\right),
\end{align}
where $\tau_x$ quantifies the horizontal gradient of the nonlinear plasma lens and we have assumed that particles are ultrarelativistic ($v\approx c$, where $c$ is the speed of light in vacuum). Using an energy-dispersed beam where $x=D_x\delta$ (to first order in $\delta$), we can compare Eqs.~\ref{eq:trajectory-evol-x} and \ref{eq:force-gradient-x} (or Eqs.~\ref{eq:trajectory-evol-y} and \ref{eq:force-gradient-y}) to find that the focusing strength of the nonlinear plasma lens is $k = k_0(1+\tau_xD_x\delta)/(1+\delta)$, which is energy-independent only when the plasma-lens nonlinearity matches the condition
\begin{align}
    \label{eq:nonlinearity-matching}
    \tau_x = \frac{1}{D_x}.
\end{align}

The focusing field in Eqs.~\ref{eq:force-gradient-x} and \ref{eq:force-gradient-y} can be achieved either via magnetic fields or via electric fields, corresponding to active or passive plasma lenses, respectively.

\begin{figure}[t]
	\centering
    \includegraphics[width=\linewidth]{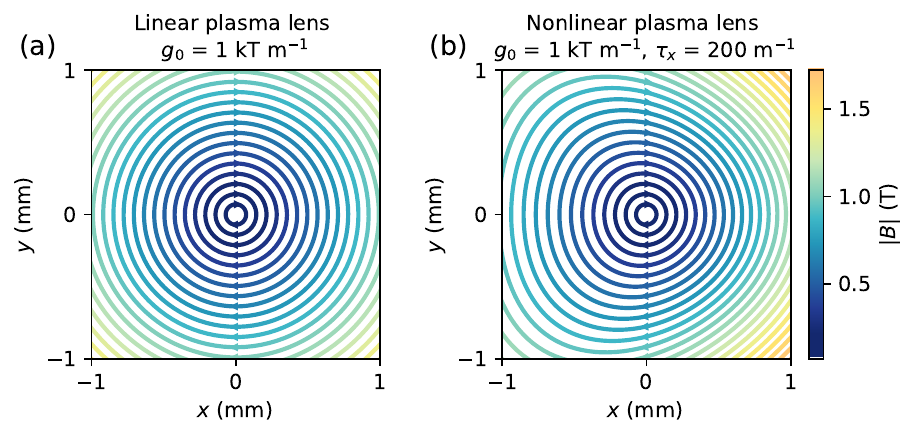}
    \caption{Comparison of the magnetic field profile in a linear (a) and nonlinear (b) active plasma lens, showing constant-field contours with the absolute strength indicated by a rainbow color map. Small arrows along the central vertical axis show the field's clockwise orientation. For increased visibility, the nonlinearity in (b) is stronger than will be typically employed.}
    \label{fig:nonlinear-plasma-lens-field}
\end{figure}

\subsubsection{Nonlinear magnetic field (active plasma lens)}
Considering first an active plasma lens, we set $\mathbf{E}=0$ and $B_z=0$ and then integrate Eqs.~\ref{eq:force-gradient-x} and \ref{eq:force-gradient-y} in both $x$ and $y$ (see the full derivation in Appendix~\ref{app:magnetic-field-profile}) to find
\begin{align}
    \label{eq:nonlinear-magnetic-field-x}
    B_x(x,y) &= - g_0 \left(y + \tau_x x y\right) + B_{x0},\\
    \label{eq:nonlinear-magnetic-field-y}
    B_y(x,y) &= g_0 \left(x +\tau_x \frac{x^2 + y^2}{2}\right) + B_{y0},
\end{align}
where $g_0 = k_0\mathcal{E}_0/qc$ is the central magnetic field gradient (i.e., at $x=0$) and $B_{x0}$ and $B_{y0}$ are constant magnetic fields in the $x$ and $y$ directions, respectively. Without loss of generality, we will set these constant fields to zero: $B_{x0} = B_{y0} = 0$. This field structure is illustrated in Fig.~\ref{fig:nonlinear-plasma-lens-field}.

From Ampere's law $\nabla\times\mathbf{B}=\mu_0\mathbf{j}$, where $\mathbf{j}$ is the current density and $\mu_0$ the vacuum permeability, we find that the active plasma lens needs to have a transversely tapered longitudinal current density
\begin{align}
    j_z(x) &= j_{z0}\left(1 + \tau_x x\right),
\end{align}
where $j_{z0} = 2g_0/\mu_0$ is the central current density. This can potentially be achieved via the Hall effect~\cite{Kunkel1981}, i.e.~by exerting a horizontal force on the charge carriers through a vertical external magnetic field $B_{y0}$~\cite{Drobniak2025a,Drobniak2025b} or via the effect of magnetization on the plasma conductivity~\cite{Davies2021}.

Although wakefield effects could be disruptive in such APLs~\cite{Pompili2018,Lindstrom2018b}, in the nonlinear APL these effects are less prominent than previously considered because the beam size is significantly larger due to chromatic dispersion from a dipole. Coulomb scattering~\cite{Montague1985,Zhao2020,Zhao2022} may also increase emittance in the APL, though it is not likely to be a major effect. Wakefield and scattering effects are briefly discussed in Sec.~\ref{sec:emittance-growth-plasma}.

\subsubsection{Nonlinear electric field (passive plasma lens)}
Considering instead a passive plasma lens, we observe from Eqs.~\ref{eq:force-gradient-x} and \ref{eq:force-gradient-y} that the transverse electric fields are equivalent to the transverse magnetic fields (i.e., $E_x\equiv-cB_y$ and $E_y\equiv cB_x$), resulting in:
\begin{align}
    \label{eq:nonlinear-electric-field-x}
    E_x(x,y) &= -g_0 c \left(x + \tau_x \frac{x^2 + y^2}{2}\right) + E_{x0}\\
    \label{eq:nonlinear-electric-field-y}
    E_y(x,y) &= -g_0 c \left(y + \tau_x x y\right) + E_{y0},
\end{align}
where $E_{x0}$ and $E_{y0}$ are constant electric fields in the $x$ and $y$ directions, respectively. Making use of Gauss' law, $\nabla\cdot\mathbf{E} = e n / \epsilon_0$, where $e$ is the electron charge, $n$ is the charge number density and $\epsilon_0$ is the vacuum permittivity, and ignoring the variation in the accelerating field (i.e., $\partial E_z/\partial z=0$), we find that the passive plasma lens must have a transverse density gradient given by
\begin{align}
    n(x) &= n_0 \left(1 + \tau_x x\right),
\end{align}
where $n_0 = -2c\epsilon_0g_0/e$. Passive plasma lenses with transverse density gradients are described in Ref.~\cite{Doss2023}. 

While achromatic optics utilizing nonlinear plasma lenses can in principle make use of either active or passive plasma lenses, in the following discussion we will assume the use of active plasma lenses as these are typically simpler to implement.


\section{Plasma-lens-based \\ achromatic staging optics}
\label{sec:achromatic-lattice}

\begin{figure*}[t]
	\centering
    \includegraphics[width=\linewidth]{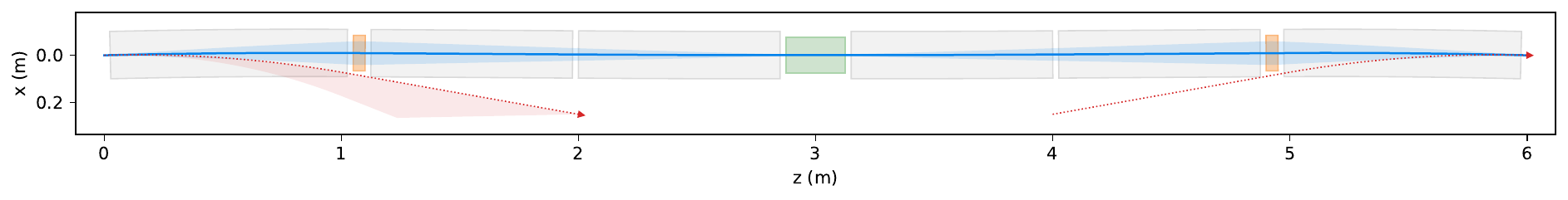}
    \caption{Top view of the achromatic staging optics based on nonlinear plasma lenses, here refocusing a \SI{10}{GeV} beam in \SI{6}{m}. The elements shown are dipoles (gray boxes), plasma lenses (orange boxes) and a sextupole (green box). The orbit (blue line) and beam-size evolution for a zero-energy-spread beam (blue area; not to scale) are indicated. Two drive beams at energy \SI{1.5}{GeV} are shown (red dotted lines): the main beam is first separated from a depleted beam driver from the previous stage (left; the red area indicates a 50\% energy loss), and then merges with a fresh beam driver for the next stage (right).}
    \label{fig:basic-setup-plasma-lenses}
\end{figure*}

The first requirement for achromatic staging between plasma stages is one or more plasma lenses, needed for refocusing the beam back to a small beta function in both the $x$ and $y$ planes. Secondly, two or more dipoles are needed to in- and out-couple drivers as well as create the exact dispersion required in the nonlinear plasma lensing; this dispersion must be canceled (to first or second order) by the end of the lattice. Thirdly, the longitudinal dispersion, $R_{56}$, must either be canceled or set to the desired value, in order to avoid bunch lengthening ($R_{56}=0$) or to enable longitudinal self-correction ($R_{56}<0$). Finally, the nonlinear focusing fields in the plasma lenses also cause emittance growth, which must to be mitigated. 

The first (and last) element of the lattice is the in-/out-coupling dipole, henceforth called the \textit{main dipole}, which has a length $L_0$ and magnetic field $B_0$. This main dipole largely defines the rest of the lattice. In the following, we will assume a nominal beam energy of $\mathcal{E}_0=\SI{10}{GeV}$ and an initial beta function of $\beta_0 = \SI{15}{mm}$, as if matched to a plasma density of \SI{5e15}{\per\cm\cubed} (or higher if density ramps are used~\cite{Marsh2005,Dornmair2015,Ariniello2019}). A main-dipole length $L_0 = \SI{1}{m}$ and field $B_0 = \SI{1}{T}$ are also assumed. Lengths of elements are chosen in order to not surpass experimentally implementable field values: i.e., \SI{\sim 1}{T} for dipoles, \SI{\sim 1}{kT/m} for plasma lenses, and \SI{\sim 10}{kT/m^2} for sextupoles. The beam and lattice parameters are summarized in Table~\ref{tab:working-point}.

\begin{table}[b]
    \centering
    \begin{tabular}{lcc}
        \textit{Beam parameters} & \textit{Symbol} & \textit{Value} \\
        \hline
        Energy & $\mathcal{E}$ & \SI{10}{GeV} \\
        Relative energy spread, rms & $\sigma_\delta$ & 2\% \\
        Charge & $Q$ & \SI{-50}{pC} \\
        Bunch length, rms & $\sigma_z$ & \SI{3}{\micro\meter} \\
        Matched beta function, $x$/$y$ & $\beta_0$ & \SI{15}{mm}\\
        Norm. emittance, $x$ & $\varepsilon_{nx}$ & \SI{10}{\milli\meter\milli\radian} \\
        Norm. emittance, $y$ & $\varepsilon_{ny}$ & \SI{0.1}{\milli\meter\milli\radian} \\
        \hline
        \cr
        \textit{Lengths} & & \\
        \hline
        Main dipole length (2$\times$) & $L_0$ & \SI{1.000}{m}\\
        Plasma-lens length (2$\times$) & $L_\mathrm{lens}$ & \SI{0.050}{m}\\
        Chicane dipole length (4$\times$) & $L_\mathrm{chic}$ & \SI{0.850}{m}\\
        Central sextupole length (1$\times$) & $L_\mathrm{sext}$ & \SI{0.250}{m}\\
        Gap between elements (10$\times$) & $\delta L$ & \SI{0.025}{m}\\
        \hline
        & \textit{Total:} & \SI{6.000}{m}\\
        \textit{Fields} &  &  \\
        \hline
        Main dipole field & $B_0$ & \SI{1.000}{T}\\
        Plasma-lens focusing gradient & $g_0$ & \SI{978.6}{T\per\meter}\\
        Plasma-lens nonlinearity & $\tau_x$ & \SI{-61.42}{\per\meter}\\
        First chicane dipole field & $B_1$ & \SI{0.022}{T}\\
        Second chicane dipole field & $B_2$ & \SI{-0.261}{T}\\
        Sextupole field & $m_\mathrm{sext}$ & \SI{9921}{T/m^2}\\
        \hline
    \end{tabular}
    \caption{Operating example used throughout this paper for achromatic staging optics with $R_{56}=0$, showing beam parameters as well as beamline-element lengths and fields. The overall bending angle of the lattice (final compared to initial trajectory) for these parameters is \SI{2.74}{\degree}.}
    \label{tab:working-point}
\end{table}

Figure~\ref{fig:basic-setup-plasma-lenses} shows the basic layout of the achromatic lattice, which is described in more detail below. Note that the nominal lattice assumes a zero longitudinal dispersion ($R_{56}=0$), but the lattice is tunable to both positive and negative $R_{56}$. 

\subsection{Satisfying the matching and cancellation requirements}

While it may be possible to satisfy many requirements using a single plasma lens, the simplest solution that satisfies all requirements is to make use of two identical plasma lenses in a mirror-symmetric lattice. The discussion above ultimately reduces to five requirements at the center or midpoint of this lattice: (i) canceling the central alpha functions
\begin{equation}
   \alpha_{x,\mathrm{mid}} = \alpha_{y,\mathrm{mid}} = 0,
\end{equation}
such that the beta functions return to the same value and the alpha functions are zero (i.e., matched to the plasma accelerator) at the end;
(ii) canceling the first-order angular dispersion in the bending plane ($x$),
\begin{equation}
    D_{x',\mathrm{mid}} = 0,
\end{equation}
such that both the first-order positional and angular dispersions are both zero at the end;
(iii) matching the longitudinal dispersion to half of its final value,
\begin{equation}
    \label{eq:R56-requirement}
    R_{56,\mathrm{mid}} = \frac{1}{2}R_{56,\mathrm{final}},
\end{equation}
such that the final longitudinal dispersion, double that of the half lattice, is $R_{56,\mathrm{final}}$ at the end, here set equal to zero; 
(iv) the first-order chromaticity in both planes should be canceled (if not at the midpoint, at least at the end)
\begin{equation}
    W_x = W_y = 0;
\end{equation}
and finally, (v) if deemed necessary due to high energy spreads or energy offsets, canceling the second-order angular dispersion
\begin{equation}
    D^{(2)}_{x',\mathrm{mid}} = 0,
\end{equation}
such that both the second-order positional and angular dispersions are both zero at the end (see Appendix~\ref{app:main-dipole-dispersions} for the definition of higher-order dispersion). The following subsections discuss how all of these requirements, in turn, can be satisfied.

\subsubsection{Matching beta functions}
Plasma accelerators provide radially symmetric focusing, which implies radially symmetric initial Twiss parameters, i.e.~$\beta_{x,0} = \beta_{y,0}$ and $\alpha_{x,0} = \alpha_{y,0}$. The radial focusing provided by plasma lenses ensures that both planes are treated identically---only one degree of freedom, and hence one plasma lens per half lattice, is therefore required. A caveat to this is that the weak focusing effect in dipoles, given by $k_\mathrm{weak} = -1/\rho^2$ where $\rho$ is the dipole bending radius, only affects the bending plane, breaking the symmetry. To account for this, a small quadrupole could be introduced to each half to add the required degree of freedom. However, for the lattices in this paper, this effect is negligible and therefore ignored. Beta-function matching can thus be performed as a one-dimensional minimization of the merit function $\alpha_{x,\mathrm{mid}}^2 + \alpha_{y,\mathrm{mid}}^2$, without regard to the strengths of the dipoles, plasma-lens nonlinearity or sextupole magnet, as none of these affect the evolution of the beta function. 

\begin{figure}[t]
	\centering
    \includegraphics[width=\linewidth]{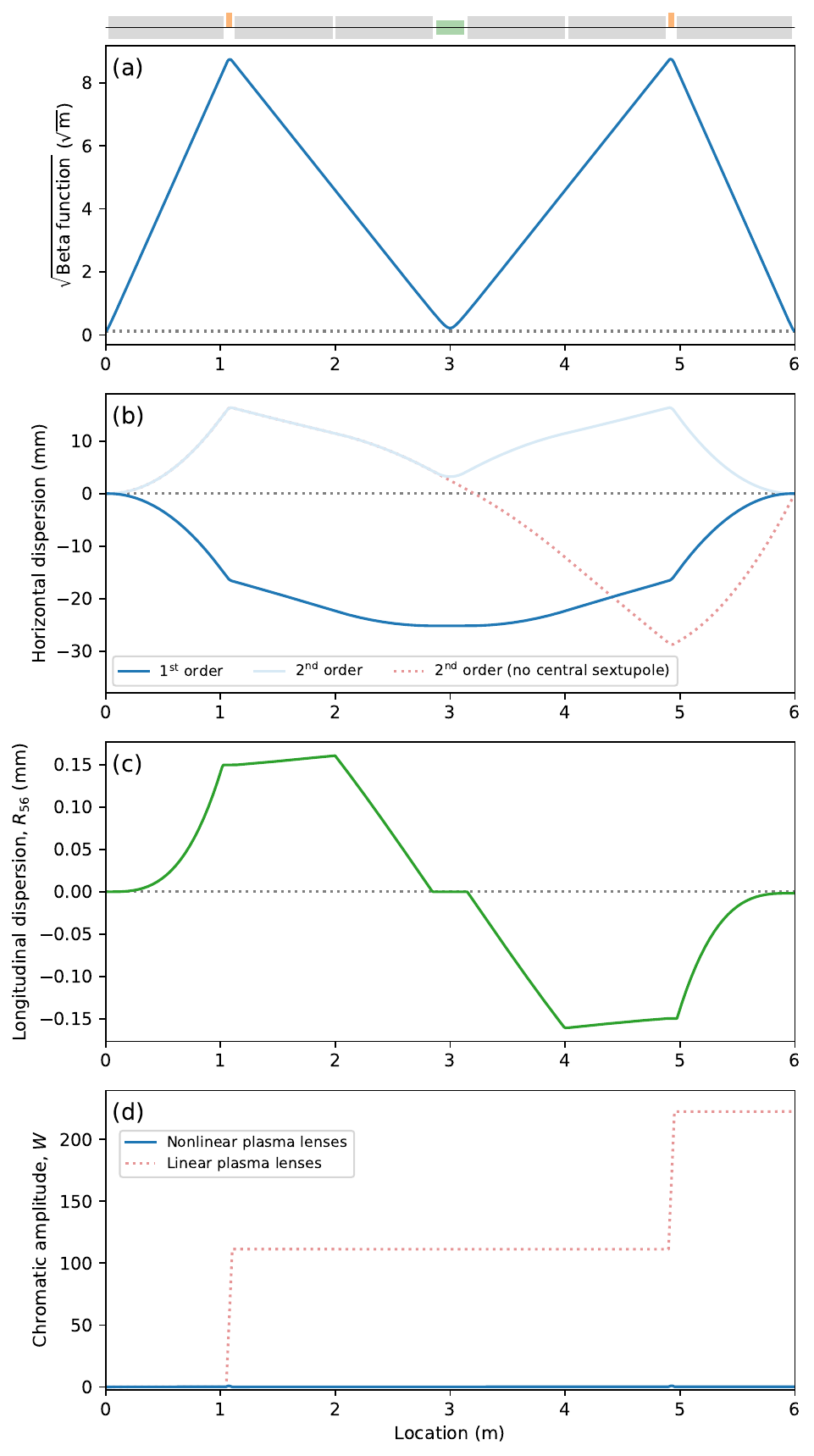}
    \caption{Achromatic staging optics at \SI{10}{GeV}, matching to and from a beta function of \SI{15}{mm}. Three elements are used (top): dipoles (gray boxes), plasma lenses (orange boxes) and a sextupole (green box). Matching or cancellation is shown for: (a) the beta function, shown by its square root (the $x$ and $y$ planes evolve identically); (b) the horizontal dispersion, shown to first and second order (dark and light blue line, respectively), as well as to second order where the central sextupole has been removed (red dotted line); (c) the longitudinal dispersion, $R_{56}$; and (d) the first-order chromaticity, shown both for nonlinear (achromatic, blue line) and linear plasma lenses (chromatic, red dotted line).}
    \label{fig:optics-functions}
\end{figure}

In principle, two matching solutions exist: one with a small central beta function, the other with a large central beta function. The former is required to give \SI{\sim180}{\degree} phase advance between the plasma lenses. This is necessary for cancellation of geometric effects from nonlinear terms (i.e., $x^2$, $y^2$ and $xy$). This \SI{180}{\degree} phase-space inversion (i.e., $x\to-x$, $x'\to-x'$, $y\to-y$ and $y'\to-y'$) is often called a $-I$ transform based on the corresponding transfer matrix. Figure~\ref{fig:optics-functions}(a) shows this solution.

Interestingly, this setup nearly perfectly preserves the beam distribution in phase space after traversing the lattice. Assuming $\alpha_{0} = 0$ (i.e., matching in the plasma accelerator), the beta function in the lens is approximately
\begin{equation}
    \beta_\mathrm{lens} \approx \frac{L^2}{\beta_0},
\end{equation}
where $L = L_0 + 2\delta L + L_\mathrm{lens}/2 \approx L_0$ is the distance to the lens ($\delta L$ is a small gap before and after the dipole and $L_\mathrm{lens}$ is the lens length), and $\beta_0 \ll L$ is the initial beta function. Since the beta function in the lenses is typically much larger than at the start and end of the lattice ($\beta_\mathrm{lens} \gg \beta_0$), and the beam goes through a waist at the midpoint, the overall phase advance $\Delta \mu = \int \beta(s)^{-1} \mathrm{d}s$ is very close to \SI{360}{\degree} (in this specific lattice, it is \SI{359.15}{\degree}). 

Finally, assuming that the plasma lenses are thin (i.e., that their length is much shorter than their focal length), we can approximate the focal length as $f_\mathrm{lens} \approx 1/(L^{-1} +l^{-1})$, where $l$ is the distance from the lens to the center of the lattice. In the example (Table~\ref{tab:working-point}), this length is just below twice the main dipole length, giving $f_\mathrm{lens} \approx 0.69 L$.

\subsubsection{Canceling first-order dispersion and matching $R_{56}$}
The first-order dispersion grows in the main dipole to become approximately (derived in Appendix \ref{app:main-dipole-dispersions})
\begin{equation}
    \label{eq:dispersion-in-lens}
    D_{x,\mathrm{lens}} \approx \frac{B_0 L_0^2 c q}{2 \mathcal{E}}
\end{equation}
at the location of the lens, where $q = \pm e$ is the particle charge (in this paper electrons are assumed; $q=-e$). The $R_{56}$ increases in the main dipole to become (see Appendix \ref{app:main-dipole-dispersions})
\begin{equation}
    \label{eq:R56-main-dipole}
    R_{56,\mathrm{lens}} \approx \frac{B_0^2 L_0^3 c^2 q^2}{6 \mathcal{E}^2}.
\end{equation}
Both first-order dispersion and $R_{56}$ are affected by dipoles, necessitating two additional dipole degrees of freedom (a \textit{central dipole chicane}) and simultaneous matching. Moreover, the plasma lens will affect the dispersion, hence there is a need to match the beta function (as described above) prior to canceling dispersion and $R_{56}$. A two-dimensional minimization is performed with the merit function $D_{x',\mathrm{mid}}^2 + (2 R_{56,\mathrm{mid}}-R_{56,\mathrm{final}})^2$, which is zero only when both $D_{x',\mathrm{mid}}=0$ and $R_{56,\mathrm{mid}}=R_{56,\mathrm{final}}/2$ (see Eq.~\ref{eq:R56-requirement}). The length of the chicane will affect the solution; in general a good solution can be found using chicane-dipole lengths between $0.5L_0$ and $L_0$ (here we choose $0.85L_0$). Figures~\ref{fig:optics-functions}(b) and (c) shows the resulting evolution of the dispersion and $R_{56}$.

\subsubsection{Canceling first-order chromaticity}
Starting from zero chromaticity, the chromatic amplitude increases in a linear-optics element approximately as $\Delta W \approx \beta_\mathrm{lens}/f_\mathrm{lens}$. If we were using linear (i.e., chromatic) plasma lenses in the lattice, this would imply a chromaticity of $\Delta W \approx (L/\beta_0)(1+L/l)$, or $1.45(L/\beta_0)$ in the given example. Since there is a \SI{180}{\degree} phase advance (i.e., a $-I$ transform) between the lenses, this implies a \SI{360}{\degree} chromatic phase advance, and hence the chromaticity adds constructively, resulting in a doubled chromatic amplitude for the full lattice of 
\begin{equation}
    \label{eq:chromaticity-plasma-lens-lattice}
    W \approx \frac{2L}{\beta_0}\left(1+\frac{L}{l}\right),
\end{equation}
as demonstrated in Fig.~\ref{fig:optics-functions}(d).

To cancel this first-order chromaticity directly in the plasma lenses, we simply follow the prescription in Eq.~\ref{eq:nonlinearity-matching} and set $\tau_x = 1/D_{x,\mathrm{lens}}$. Figure~\ref{fig:optics-functions}(d) shows that in this case, no first-order chromaticity is introduced at any point; the chromaticity correction is thus \textit{fully local}~\cite{Raimondi2001}. 

\subsubsection{Canceling second-order dispersion}
Finally, the second-order dispersion is affected by both dipoles and plasma lenses; here also the nonlinearity $\tau_x$ matters. While the second-order dispersion does not need to be canceled if the energy spread is small, it may be required for energy spreads above 1\% rms. To cancel the second-order dispersion, we wish to use a degree of freedom that does not affect any of the above matching (i.e., of the beta function, first-order dispersion and $R_{56}$). A sextupole is therefore the element of choice.

In order to avoid affecting chromaticity when introducing a sextupole, we must locate this sextupole at the center of the lattice. Here, the dispersion is large and the beam size small (and going through a waist), which ensures that only the second-order dispersion is affected, with negligible effect on the chromaticity [as seen in Fig.~\ref{fig:optics-functions}(d)]. The second-order dispersion can therefore be canceled independently with a one-dimensional minimization with the merit function $(D^{(2)}_{x',\mathrm{mid}})^2$. The resulting evolution is shown in Fig.~\ref{fig:optics-functions}(b), which also shows that, with no sextupole, the second-order positional dispersion is nearly canceled but the second-order angular dispersion (i.e., the longitudinal derivative of the second-order dispersion) is not; this means that, unless mitigated, as particles are offset in energy, their angle increases with the square of the relative energy offset (i.e., a curved distribution in $x'$--$\delta$ space).

\subsection{Driver separation}
\label{sec:driver-separation}
In principle, the achromatic lattice is agnostic to the type of driver used; the main dipole can be used to separate the transported beam from both a laser driver and a particle-beam driver. However, there are practical limitations in each case that need to be considered.

For a laser driver, the key consideration is whether the laser, which continues on a straight trajectory while the beam is being bent, can pass outside the plasma lens. When only traversing dipoles and drift spaces, the beam's separation from the laser axis is given by, to first order (i.e., assuming small angles), the first-order dispersion (see Eq.~\ref{eq:dispersion-in-lens}). Hence the separation at the lens is
\begin{equation}
    \Delta x_\mathrm{laser} \approx \frac{B_0 L_0^2cq}{2\mathcal{E}} = D_{x,\mathrm{lens}}.
\end{equation}
or about \SI{15}{mm} for the \SI{10}{GeV} example above (see Table~\ref{tab:working-point}). The beam passes within the plasma lens, which typically has a transverse extent of 1--\SI{3}{mm}~\cite{vanTilborg2015,Pompili2018,Lindstrom2018a}, leaving a maximum of \SI{12}{mm} for the laser to pass at a distance of \SI{1}{m} from the stage. This corresponds to a maximum divergence angle of \SI{12}{mrad} or an f-number of $f/42$.

For a beam driver, say with a nominal energy of $\mathcal{E}_\mathrm{d}$, the separation to the transported beam is given by
\begin{equation}
    \label{eq:beam-driver-separation}
    \Delta x_\mathrm{beams} \approx \frac{B_0 L_0^2cq}{2}\left(\frac{1}{\mathcal{E}_\mathrm{d}}-\frac{1}{\mathcal{E}}\right) = D_{x,\mathrm{lens}}\left(\frac{\mathcal{E}}{\mathcal{E}_\mathrm{d}}-1\right).
\end{equation}
Since a depleted beam driver will have up to 100\% energy spread, it must have a lower nominal energy than the nominal lattice energy (i.e., $\mathcal{E}_\mathrm{d}<\mathcal{E}$) to avoid overlap. Moreover, Eq.~\ref{eq:beam-driver-separation} implies a maximum allowable driver energy: assuming a \SI{3}{mm} minimum separation, the maximum driver energy in the example is $\mathcal{E}(\Delta x_\mathrm{beams}/D_{x,\mathrm{lens}}+1)^{-1}$ or \SI{8.3}{GeV}. Note also that when the driver energy is significantly lower than the lattice energy (as illustrated in Fig.~\ref{fig:basic-setup-plasma-lenses}), the separation becomes very large; this is not the case for laser drivers. 

\subsection{Particle-tracking simulation}
\label{sec:tracking-simulation}

\begin{figure*}[t]
	\centering
    \includegraphics[width=\linewidth]{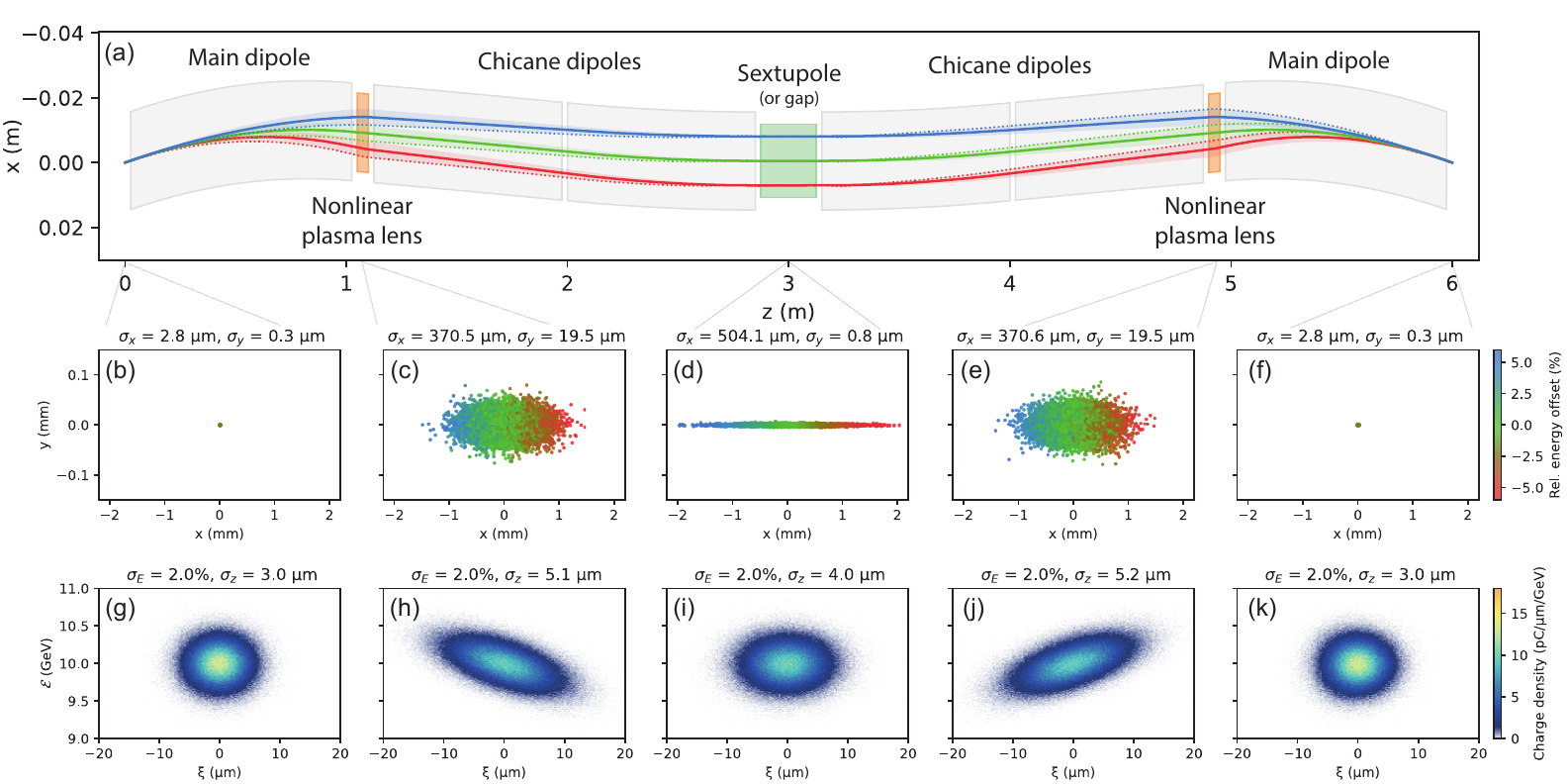}
    \caption{(a) Working principle of the achromatic staging optics, showing the elements, orbit (solid lines) and beam-size variation (shaded areas) of three individual energy slices (here using an exaggerated energy spread of $\pm$30\%); low energy (red), nominal energy (green) and high energy (blue). Also represented are single particles with an initial angle (dotted lines). The beam is initially chromatically dispersed by a dipole, achromatically focused by the first nonlinear plasma lens, then refocused at the midpoint, where a sextupole corrects second-order dispersion, next refocused by the second plasma lens, and finally undispersed by the final dipole. The chicane dipoles are used to control the $R_{56}$. (b--f) The evolution of the transverse $x$--$y$ profile of the beam particles is shown for 5 locations along the lattice, where the color bar (red--green--blue) indicates relative energy offset. At the midpoint (d) the beam is both tightly focused and highly dispersed in $x$; ideally suited for a spectrum measurement. (g--k) The evolution is also shown in the longitudinal $\xi$--$\mathcal{E}$ phase space. Note that the bunch length is larger at the midpoint compared to the start (i); this is not due to $R_{56}$ but instead due to $R_{52}$, which is canceled at the end of the lattice.}
    \label{fig:phase-space-evolution}
\end{figure*}

In order to understand the transverse and longitudinal dynamics of the staging lattice, and to demonstrate beam-quality preservation, it is necessary to perform particle tracking. Simulations are performed with the \textsc{ABEL} framework~\cite{Chen2025} using the tracking code \textsc{ImpactX}~\cite{Huebl2022}, which implements regular accelerator elements (i.e., dipoles, quadrupoles, sextupoles, etc.) as well as the nonlinear plasma lens introduced here. \textsc{ImpactX} includes effects such as coherent and incoherent synchrotron radiation, which is turned on in all simulations unless otherwise stated. An implementation was also made in \textsc{ELEGANT}, giving similar results. The beam parameters are given in Table~\ref{tab:working-point}. Simulations were performed with 100,000 or more macro-particles, employing \textsc{ImpactX}'s exact tracking model for all elements, each of which were split into 50 longitudinal slices. All scripts used in this paper have been made publicly available \cite{Zenodo}.

\subsubsection{Visualizing the evolution in 6D phase space}
The complex and interwoven operation of the achromatic lattice is not straightforward to understand. For pedagogical purposes, therefore, we have included Fig.~\ref{fig:phase-space-evolution} to visualize the evolution of the beam throughout the lattice. The beam orbit and size is shown for three different energy slices, and the distribution of tracked particles is shown in both the $x$--$y$ transverse projection and in the $\xi$--$\mathcal{E}$ longitudinal phase space, where $\xi=z-ct$ is the co-moving coordinate ($z$ is the longitudinal position and $t$ is time). The simulation demonstrates how the beam is transported achromatically by the lattice to recreate its initial particle distribution.

\subsubsection{Preservation of emittance and other beam qualities}
\begin{figure}[t]
	\centering
    \includegraphics[width=\linewidth]{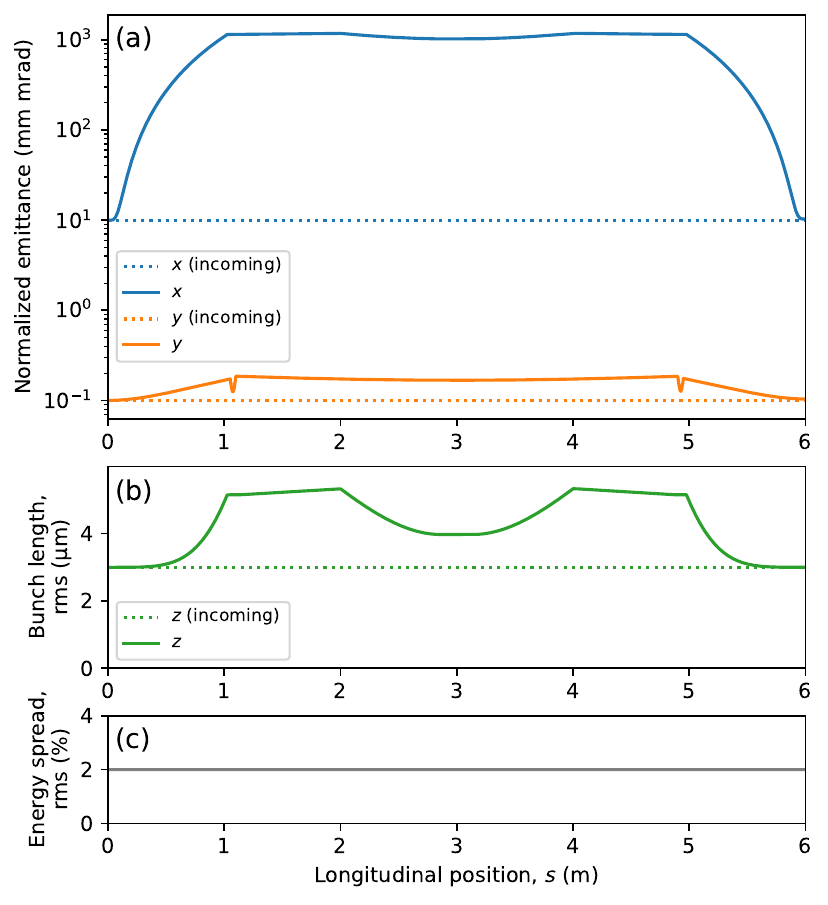}
    \caption{Simulated evolution of beam qualities for the staging optics at \SI{10}{GeV} (see Table~\ref{tab:working-point} for the full parameter set) as a function of the longitudinal position in the lattice, demonstrating preservation of: (a) the normalized emittance in both the horizontal (blue) and vertical (orange) planes; (b) the bunch length; and (c) the energy spread.}
    \label{fig:beam-quality-preservation}
\end{figure}

The ultimate goal of the staging optics is to transport beams from a plasma accelerator without degrading beam qualities. Figure~\ref{fig:beam-quality-preservation} shows the evolution of the transverse emittances, bunch length and energy spread. 

The horizontal emittance increases dramatically due to the introduction of a large horizontal dispersion [see Fig.~\ref{fig:optics-functions}(b)], but returns to its initial value at the end. The vertical emittance temporarily increases by a factor of 2, from a combination of the intrinsic chromatic emittance growth of a dipole/drift as well as a small geometric aberration (both which are later canceled). 

The bunch length also increases temporarily before returning to its initial value, for two reasons: firstly, due to the $R_{56}$ present in the lattice before its final cancellation [see Fig.~\ref{fig:optics-functions}(c)]; and secondly, due to the $R_{52}$ (i.e., correlation between horizontal angle and longitudinal position) also present before its cancellation at the end of the lattice. While the former ($R_{56}$) is canceled at the midpoint of the lattice, the latter ($R_{52}$) is not, leading to a lengthening of the bunch at this point [see Figs.~\ref{fig:phase-space-evolution}(i) and \ref{fig:beam-quality-preservation}(b)]. The $R_{52}$ is canceled at the end because of the $-I$ transform between the two plasma lenses, implying that particles taking the ``inner route'' in the first half must take the ``outer route'' in the second half and vice versa [as illustrated by the dotted lines in Fig.~\ref{fig:phase-space-evolution}(a)]. 

Finally, none of the elements a priori affect the energy of the particles, hence the energy spread is preserved. That said, synchrotron radiation, chiefly from the dipoles, can affect the energy and energy spread, but this effect is found to be negligible for the example parameters shown in Table~\ref{tab:working-point}. The effect of synchrotron radiation does, however, become important at much higher and at much lower energies, as discussed in Sec.~\ref{sec:energy-scaling-and-sr}.

Figure~\ref{fig:beam-quality-preservation} demonstrates preservation of the horizontal and vertical emittance (to within 3\%), as well as preservation of the bunch length (to within 0.1\%) and energy spread (to within 0.01\%), for a \SI{10}{GeV} beam with an energy spread of 2\% rms. The full 6D phase space of the beam is therefore preserved.

\subsubsection{Tunable $R_{56}$}

\begin{figure}[b]
	\centering
    \includegraphics[width=\linewidth]{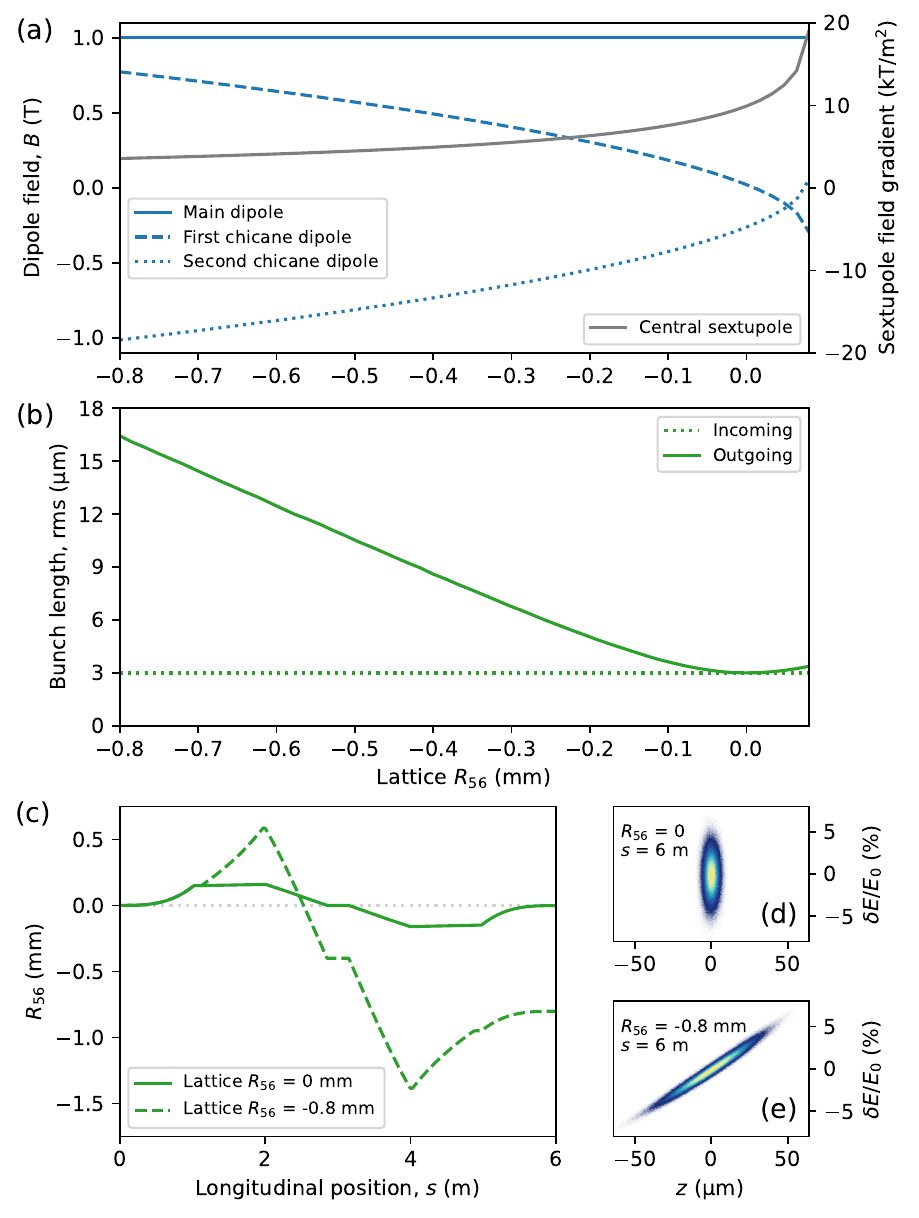}
    \caption{(a) The $R_{56}$ of the lattice is tunable by adjusting the two chicane dipoles (dashed and dotted blue lines). The central-sextupole strength (gray line) is adjusted accordingly. The field in the main dipoles (solid blue line) and the plasma lenses (not shown) remain constant. Since the two chicane dipoles vary oppositely, the overall lattice bending angle is constant. (b) Starting from an initially unchirped bunch (dotted green line), a variation in bunch length with different $R_{56}$ (solid green line) results. (c) The intermediate evolution of the $R_{56}$ is shown for two specific examples: \SI{0}{mm} (solid green line) and \SI{-0.8}{mm} (dashed green line), with corresponding longitudinal phase-space distributions shown in (d) and (e).}
    \label{fig:tunable-R56}
\end{figure}

While the standard solution presented in this paper has zero $R_{56}$, it is a key feature that the lattice should have a tunable $R_{56}$, in particular to negative values as required for the longitudinal self-correction mechanism~\cite{Lindstrom2021b}. Figure~\ref{fig:tunable-R56} highlights this capability, showing that for the staging lattice presented, longitudinal dispersions from \SI{-0.8}{mm} to \SI{+0.08}{mm} are attainable without exceeding the field of the main dipole (here \SI{\pm 1}{T}). A change in $R_{56}$ only requires adjustment of the chicane dipoles and the central sextupole, leaving the fields in the main dipoles and plasma lenses unchanged.

\section{For comparison: quadrupole-based staging optics}
\label{sec:achromatic-quad-lattice}

\begin{figure}[h]
	\centering
    \includegraphics[width=\linewidth]{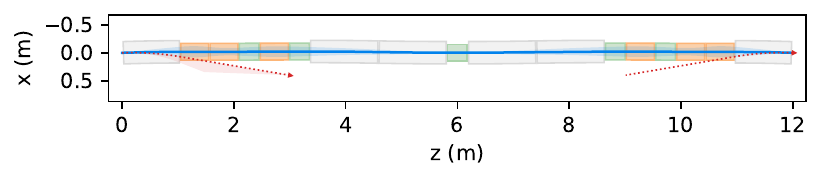}
    \caption{Top view of the quadrupole- and sextupole-based achromatic staging optics at \SI{10}{GeV}, with colors as in Fig.~\ref{fig:basic-setup-plasma-lenses} except that plasma lenses have been exchanged for quadrupoles (orange boxes). The lattice is \SI{12}{m} long and uses 17 magnets; 6 dipoles, 6 quadrupoles and 5 sextupoles.}
    \label{fig:basic-setup-quads}
\end{figure}

\begin{figure}[b]
	\centering
    \includegraphics[width=\linewidth]{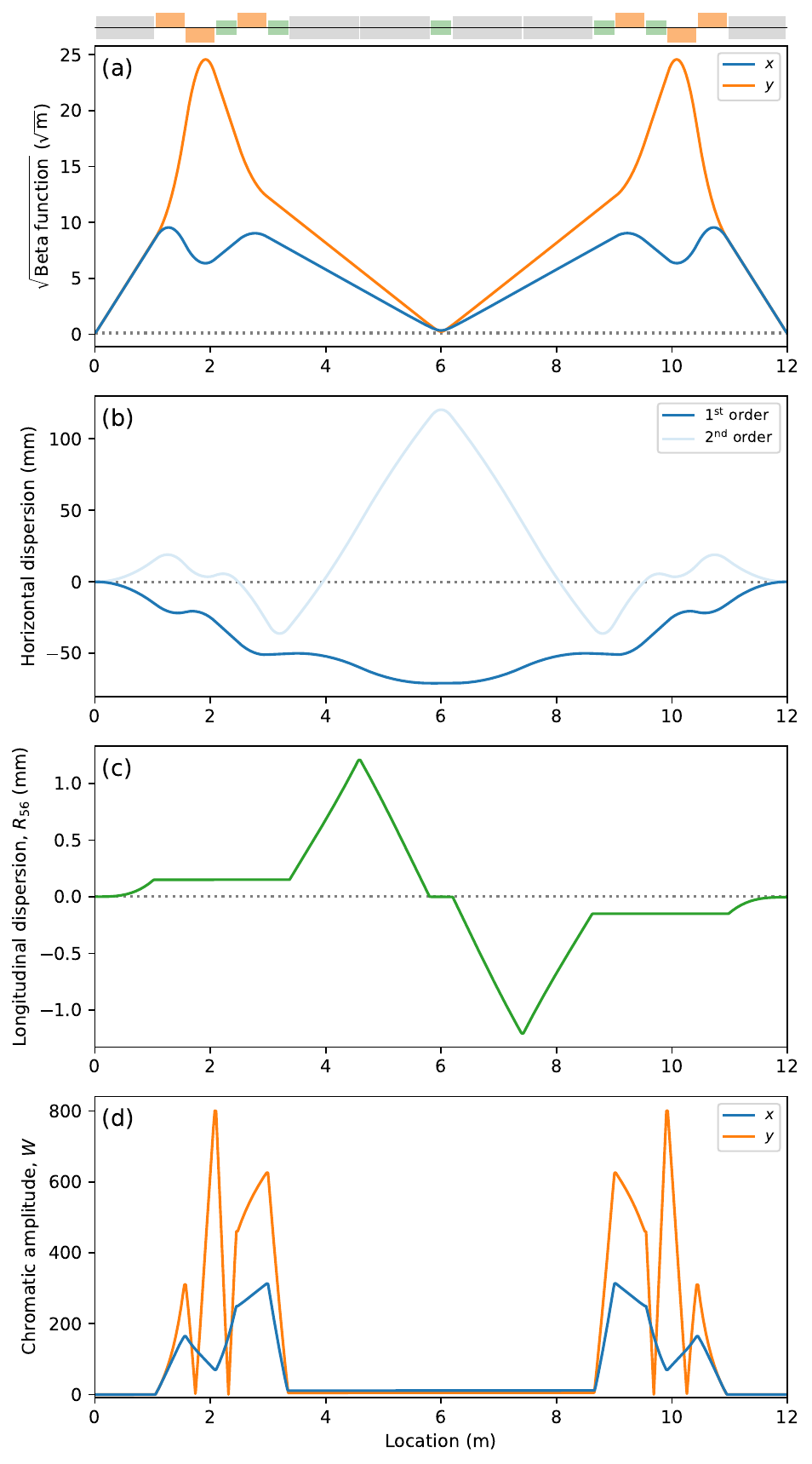}
    \caption{Achromatic staging optics using quadrupoles and sextupoles, at 10 GeV. The plots show the evolution of: (a) beta functions in the horizontal/vertical plane (blue/orange lines, respectively); (b) the first- and second-order horizontal dispersion (blue and light blue, respectively); (c) $R_{56}$; and (d) the chromatic amplitude in both planes. Here, the chromatic amplitude is not corrected fully locally, but only at the end of each quadrupole triplet.}
    \label{fig:optics-functions-quads}
\end{figure}

The merits of a plasma-lens-based staging lattice can be seen by comparison to a more conventional quadrupole- and sextupole-based staging lattice. Such lattices have previously been proposed, e.g.~in Refs.~\cite{Chance2014,Lindstrom2016a,Lindstrom2016b,Lindstrom2019}. However, by applying similar steps as in Sec.~\ref{sec:achromatic-lattice} and fully exploiting the concept of local chromaticity correction, as in modern linear-collider final-focus systems, we can further optimize such quadrupole-based lattices. The resulting lattice is shown in Figs.~\ref{fig:basic-setup-quads} and \ref{fig:optics-functions-quads}, with elements specified in full in Table~\ref{tab:working-point-quads}.

\begin{table}[b]
    \centering
    \begin{tabular}{lcc}
        \textit{Lengths} & \textit{Symbol} & \textit{Value} \\
        \hline
        Main dipole length (2$\times$) & $L_0$ & \SI{1.000}{m}\\
        Quadrupole length (6$\times$) & $L_\mathrm{quad}$ & \SI{0.500}{m}\\
        Sextupole length (5$\times$) & $L_\mathrm{sext}$ & \SI{0.350}{m}\\
        Chicane dipole length (4$\times$) & $L_\mathrm{chic}$ & \SI{1.200}{m}\\
        Gap between elements (18$\times$) & $\delta L$ & \SI{0.025}{m}\\
        \hline
        & \textit{Total:} & \SI{12.000}{m}\\
        \textit{Fields} &   &  \\
        \hline
        Main dipole field & $B_0$ & \SI{1.000}{T}\\
        First quadrupole field gradient& $g_1$ & \SI{130.5}{T/m}\\
        Second quadrupole field gradient & $g_2$ & \SI{-138.2}{T/m}\\
        First sextupole field gradient & $m_1$ & \SI{8555}{T/m^2}\\
        Third quadrupole field gradient & $g_3$ & \SI{53.8}{T/m}\\
        Second sextupole field gradient & $m_2$ & \SI{-8838}{T/m^2}\\
        First chicane dipole field & $B_1$ & \SI{0.558}{T}\\
        Second chicane dipole field & $B_2$ & \SI{-0.500}{T}\\
        Central sextupole field gradient & $m_3$ & \SI{3909}{T/m^2}\\
        \hline
    \end{tabular}
    \caption{Beamline-element lengths and fields for a quadrupole- and sextupole-based achromatic lattice with $R_{56}=0$, at energy \SI{10}{GeV} and matched beta function \SI{15}{mm}. The overall bending angle of this lattice is \SI{3.67}{\degree}.}
    \label{tab:working-point-quads}
\end{table}

The design process can be summarized as follows. Just as in the plasma-lens-based lattice, mirror symmetry and a $-I$ transform between the two ``lenses", now groups of focusing elements, are required to cancel geometric effects. The plasma lenses are replaced by triplets of quadrupole magnets, whose strengths are matched so as to produce $\alpha_x = \alpha_y = 0$ at the lattice center, with the remaining quadrupole degree of freedom used to control the ratio $\beta_x/\beta_y$ (here set to 2) at the lattice center. Next, the first-order horizontal dispersion and $R_{56}$ are canceled using the two dipole degrees of freedom of the central magnetic chicane, and a central sextupole is used to cancel the second-order dispersion---all just as in the plasma-lens-based lattice. Finally, the first-order chromatic amplitude is canceled semi-locally in the $x$ and $y$ planes using two sextupoles placed close to the quadrupole triplet. To function properly, these sextupoles must be located at positions that have significantly different $\beta_x/\beta_y$ ratios; this implies that the two sextupoles should not be placed symmetrically within the quadrupole triplet, but instead one inside and one outside the triplet. 

In principle, two distinct solutions exists: one where the first quadrupole defocuses in the $y$-plane, and another where it defocuses in the $x$-plane. While both are viable solutions, the former is chosen, as it leads to a smaller buildup of first- and second-order dispersion and hence needs weaker chicane dipole strengths. 

The lengths of each element are chosen such that the field values are experimentally attainable (i.e., maximum \SI{\sim 150}{T/m} for quadrupoles and \SI{\sim 10000}{T/m^2} for sextupoles). The total length of the quadrupole-based lattice is found to be double that of the plasma-lens-based lattice (\SI{12}{m} versus \SI{6}{m}, respectively, at \SI{10}{GeV}).

\begin{figure}[t]
	\centering
    \includegraphics[width=\linewidth]{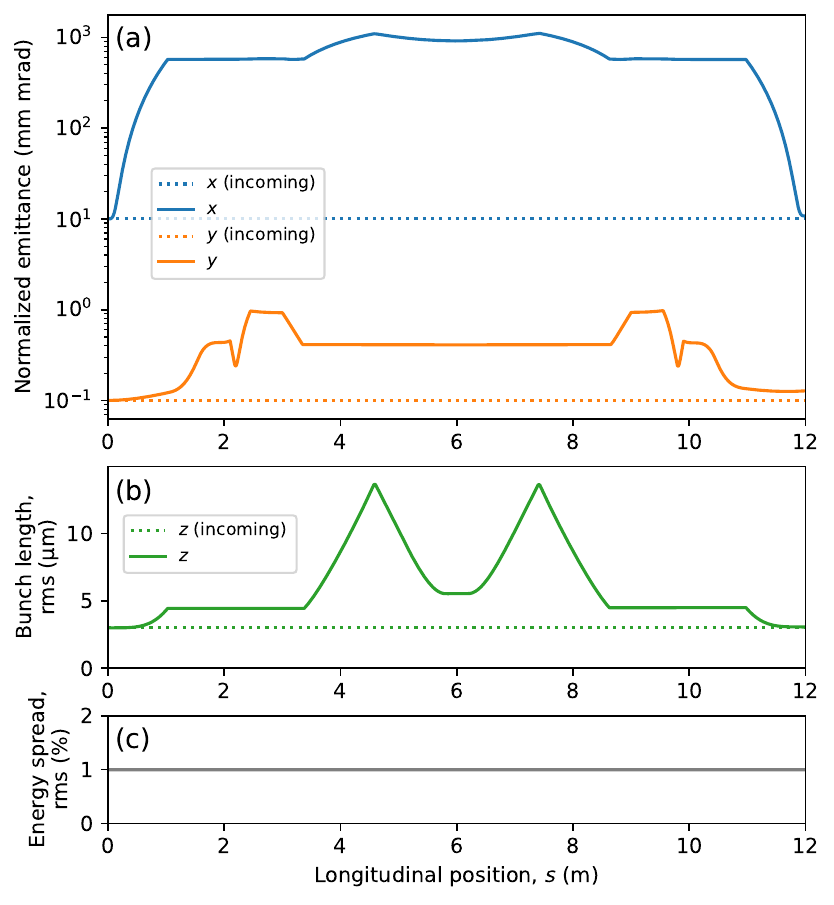}
    \caption{Evolution of beam qualities in the quadrupole- and sextupole-based optics at \SI{10}{GeV}, demonstrating preservation of: (a) the normalized emittance in both the horizontal (blue) and vertical (orange) planes; (b) the bunch length; and (c) the energy spread. Note that this simulation was performed with a relative energy spread of 1\% rms, as the 2\% rms used in the plasma-lens-based lattice would not preserve the emittance.}
    \label{fig:beam-quality-preservation-quads}
\end{figure}

A comparison of Figs.~\ref{fig:optics-functions} and \ref{fig:optics-functions-quads} shows that in the quadrupole-based lattice: the beta function increases significantly more in the $y$-plane; both the first- and second-order dispersion as well as the $R_{56}$ grow comparatively larger before being canceled; and that the first-order chromatic amplitude grows significantly within each triplet (by an order of magnitude more than in a linear plasma lens) before being canceled by the sextupoles. While the optics functions shown are all canceled at the end of the lattice, the larger intermediate values are associated with larger higher-order effects (e.g., third-order dispersion, second-order chromaticity)---these are less problematic in the nonlinear plasma lenses due to their fully local chromaticity correction. The performance of the quadrupole-based lattice, in terms of energy bandwidth and geometric aberrations, will therefore be worse than that of the plasma-lens-based lattice. Figure~\ref{fig:beam-quality-preservation-quads} shows the evolution of beam qualities in the lattice, preserving emittances (except for a small increase in $y$ from geometric effects), bunch length and energy spread, though only for a smaller energy spread of 1\% rms---the highest energy spread for which emittance is preserved in this lattice.


\section{Performance and limitations}
\label{sec:performance-and-limitations}

In Sec.~\ref{sec:achromatic-lattice}, we established that the achromatic lattice preserves all beam qualities for the example parameters given. However, if such a lattice is to be applied across a wide range of different plasma accelerators, it is necessary to understand under what specific conditions the lattice preserves emittance, and when it does not. Below we attempt to analytically model the key sources of emittance growth---chromaticity (Sec.~\ref{sec:chromatic-aberrations}), geometric effects (Sec.~\ref{sec:geometric-effects}), misalignments (Sec.~\ref{sec:tolerances}) and plasma-specific effects (Sec.~\ref{sec:emittance-growth-plasma})---and then test the validity of the models using numerical simulations. Synchrotron radiation is discussed in a separate section below (Sec.~\ref{sec:energy-scaling-and-sr}).

\subsection{Chromatic aberrations from large energy spreads and energy offsets}
\label{sec:chromatic-aberrations}

\begin{figure}[t]
	\centering
    \includegraphics[width=\linewidth]{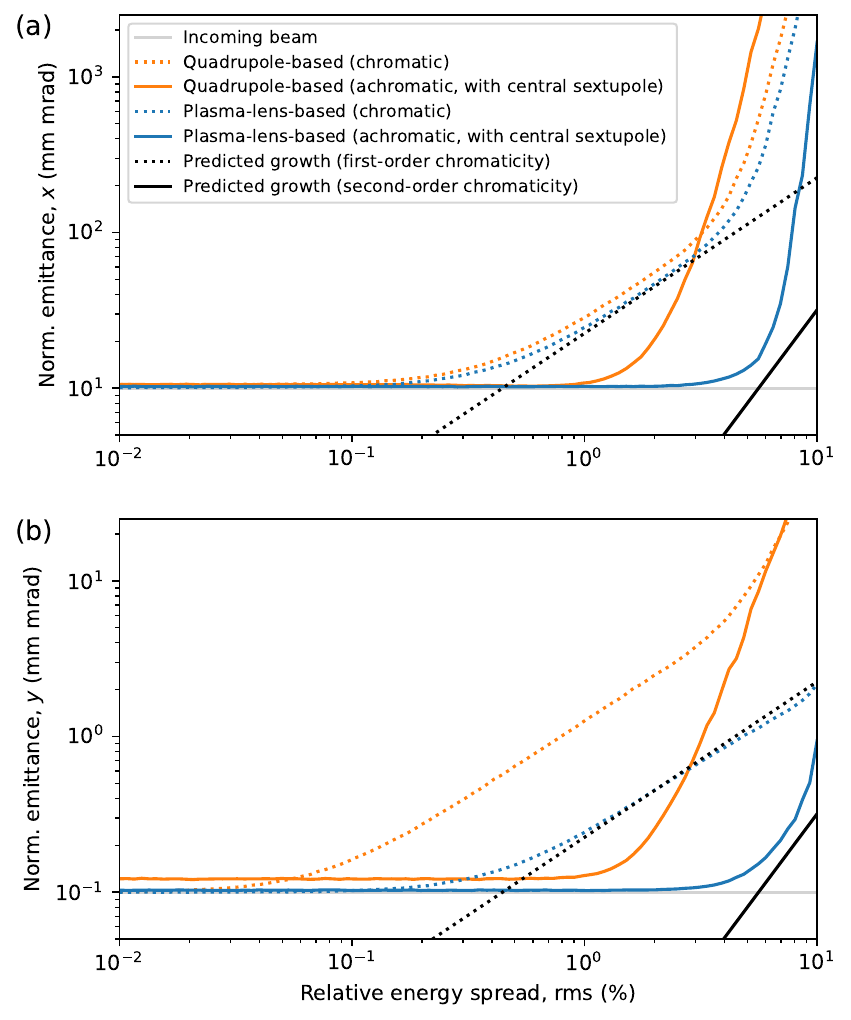}
    \caption{Emittance growth due to chromaticity in the horizontal (a) and vertical planes (b), shown with a scan of the energy spread. The beam energy is \SI{10}{GeV} and the matched beta function is \SI{15}{mm}. Lattices based on plasma lenses (blue) and quadrupoles (orange) are compared, each showing chromatic/linear optics (dotted lines) and achromatic/nonlinear optics (solid lines). Analytic predictions for the plasma-lens-based lattice are shown, to first (black dotted line; Eq.~\ref{eq:first-order-chromaticity}) and second order in chromaticity (black solid line; Eq.~\ref{eq:emittance-growth-second-order-chromaticity}). The initial emittance is also indicated (gray line).}
    \label{fig:chromaticity} 
\end{figure}

The proposed lattice is achromatic within a certain range of particle energies. Figure~\ref{fig:chromaticity} shows the transverse emittance growth in the horizontal and vertical planes for the beam parameters listed in Table~\ref{tab:working-point} but exploring a range of rms energy spreads. It indicates that the plasma-lens-based achromatic lattice preserves emittance up to an energy spread of 3--5\% rms beyond which the emittance rapidly increases with energy spread. The quadrupole-based achromatic lattice behaves similarly but performs somewhat worse, preserving emittance up to about 1\% rms energy spread. Without chromatic correction, emittance is preserved for the plasma-lens-based lattice up to 0.1--0.2\% rms energy spread, while for the quadrupole-based lattice this degrades to 0.05\% in its defocusing plane, $y$, which has higher chromaticity. 

For the chromatic lattices using linear optics only, the emittance growth is linear with energy spread (derived by combining Eqs.~\ref{eq:chromaticity} and \ref{eq:chromaticity-plasma-lens-lattice}),
\begin{equation}
    \label{eq:first-order-chromaticity}
    \frac{\Delta\varepsilon _{nx}}{\varepsilon _{nx}} \approx 2\left(1+\frac{L}{l}\right)\frac{L}{\beta_0}\sigma_\delta.
\end{equation}
Here, $L \approx L_0$ is the distance to the middle of the first lens and $l = 2L_\mathrm{chic}+L_\mathrm{sext}/2 + L_\mathrm{lens}/2 +3\delta L$ is the distance from the first lens to the lattice center ($L_\mathrm{chic}$ is the chicane-dipole length, $L_\mathrm{sext}$ is the central sextupole length, $L_\mathrm{lens}$ is the lens length and $\delta L$ is the gap between elements, as defined in Table~\ref{tab:working-point}). 
However, the achromatic lattice cancels first-order chromaticity such that only quadratic energy terms or higher remain. The relative emittance growth (see Appendix~\ref{app:second-order-chromaticity} for the derivation) is then given by
\begin{equation}
    \label{eq:emittance-growth-second-order-chromaticity}
    \frac{\Delta\varepsilon _{nx}}{\varepsilon _{nx}} = \frac{\Delta\varepsilon _{ny}}{\varepsilon _{ny}} \approx 2\left(1+\frac{L}{l}\right) \sigma_{\delta}^2 \sqrt{ 2 \frac{L^2}{\beta_0^2} + 3\left(\frac{l}{L}+1\right)^2}. 
\end{equation}
This can be recast as an effective maximum bandwidth, defined as what can be transported with less than 41\% emittance growth (i.e., $\Delta\varepsilon _{n}/\varepsilon _{n}=1$ added in quadrature to the initial emittance):
\begin{equation}
    \sigma_{\delta}^{\max} \lesssim \frac{1}{\sqrt{2\left(1+\frac{L}{l}\right) \sqrt{ 2 \frac{L^2}{\beta_0^2} + 3\left(\frac{l}{L}+1\right)^2}}}. 
\end{equation}
Assuming similar lattice and beam parameters as the working point in Table~\ref{tab:working-point}, where $l/L \approx 2$ and $\beta_0 \ll L$, this energy-spread limit simplifies to approximately 
\begin{equation}
    \sigma_{\delta}^{\max} \lesssim \frac{1}{2}\sqrt{\frac{\beta_0}{L}}.
\end{equation}
This can be compared to the equivalent limit from first-order chromaticity (using Eq.~\ref{eq:first-order-chromaticity}), which would be $\sigma_{\delta}^{\max} \lesssim \beta_0/3L$, again assuming $l/L \approx 2$. For parameters in Table~\ref{tab:working-point}, the first- and second-order limits correspond to about 0.5\% and 6\% rms, respectively.

\begin{figure*}[t]
	\centering
    \includegraphics[width=\linewidth]{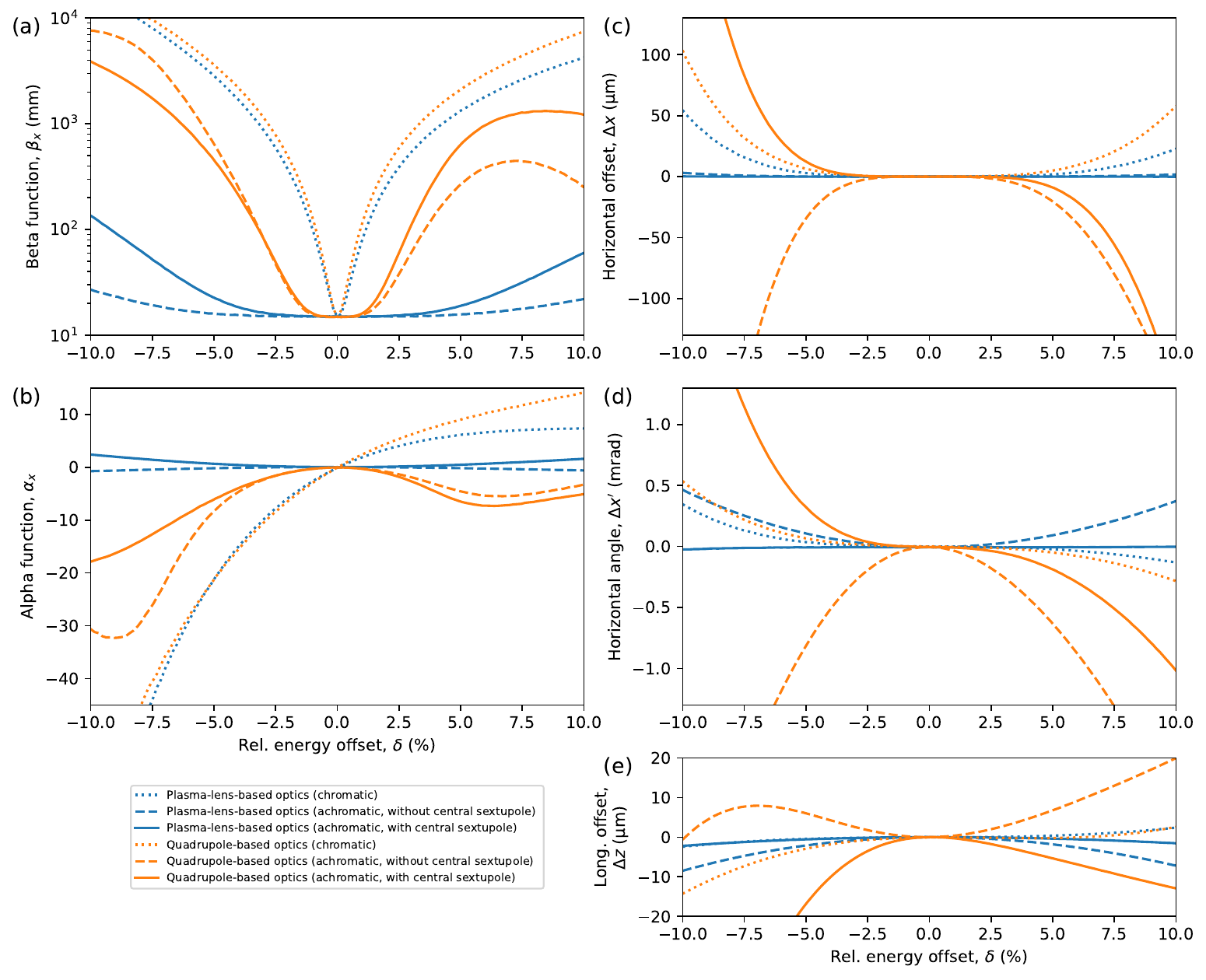}
    \caption{Effect of an energy offset on Twiss parameters (a) $\beta$ and (b) $\alpha$, as well as centroid offsets in (c) $x$, (d) $x'$ and (e) $z$, for a \SI{10}{GeV} bunch with zero energy spread. Both plasma-lens- and quadrupole-based solutions (blue and orange lines, respectively) are shown, each using either no chromatic correction (dotted lines), chromatic correction without a central sextupole (dashed lines) or with a central sextupole (solid lines). Generally, the plasma-lens-based solution performs better, and the central sextupole trades better performance in offsets (i.e., second-order dispersion) for worse performance in Twiss.}
    \label{fig:energy-offsets}
\end{figure*}

Chromatic aberrations encompass both the effect of the envelope (i.e., Twiss parameters) as well as the beam centroid. To understand the chromatic effects in more detail, we can look at the effect of an energy offset (assuming zero energy spread), as shown in Fig.~\ref{fig:energy-offsets}. In particular, here we observe that the addition of the central sextupole reduces the centroid offsets for large energy offsets (its purpose is to cancel second-order dispersion), but this comes at the cost of an increased second-order chromaticity in the Twiss parameters ($\beta$ and $\alpha$).

\subsection{Geometric aberrations from large emittances and small matched beta functions}
\label{sec:geometric-effects}

Local chromaticity correction, as utilized in the proposed lattice, introduces nonlinear focusing fields, i.e.~terms including $x^2$, $y^2$ and $xy$. These terms are negligible for sufficiently small beam sizes, but will introduce emittance growth for large beam sizes. While this emittance growth is cancelled to first order in nonlinearity [i.e., $\mathcal{O}(\tau_x)$] by the use of a $-I$ transform (\SI{180}{\degree} phase advance) between the plasma lenses, higher order effects will appear for sufficiently large beam sizes.

In order to estimate this emittance growth, we can consider a simplified lattice with thin plasma lenses and a single particle with nominal energy and offsets $x_0$ and $y_0$ and angles $x'_0$ and $y'_0$. Transporting this particle through the drifts and applying the nonlinear kicks in the plasma lenses, and finally averaging over a Gaussian beam distribution in phase space (see Appendix~\ref{app:nonlinear-forces} for the full derivation), we estimate relative emittance growths
\begin{eqnarray}
    \label{eq:geometric-aberration-x}
    \frac{\Delta\varepsilon _{nx}}{\varepsilon _{nx}} &\approx& \frac{\tau_x^2 L^3}{\beta_0^2 \gamma} \left(1+\frac{L}{l}\right)\left(\frac{l}{L}+1\right)\sqrt{6\varepsilon _{nx}^2 + 18\varepsilon _{ny}^2} \\
    \label{eq:geometric-aberration-y}
    \frac{\Delta\varepsilon _{ny}}{\varepsilon _{ny}} &\approx& \frac{\tau_x^2 L^3}{\beta_0^2 \gamma} \left(1+\frac{L}{l}\right)\left(\frac{l}{L}+1\right)\sqrt{18\varepsilon _{nx}^2 + 6\varepsilon _{ny}^2}.
\end{eqnarray}
Equations \ref{eq:geometric-aberration-x} and \ref{eq:geometric-aberration-y} only retain the lowest remaining order in nonlinearity [i.e., $\mathcal{O}(\tau_x^2)$], neglecting higher orders. These can be recast into an upper limit on emittance, again defined as that transportable with less than 41\% emittance growth (i.e., $\Delta\varepsilon _{n}/\varepsilon _{n}=1$) in either plane:
\begin{equation}
    \label{eq:emittance-limit-geometric-effects}
    \varepsilon ^{\max}_n \lesssim \frac{\beta_0^2 \gamma}{\sqrt{18}\left(1+\frac{L}{l}\right)\left(\frac{l}{L}+1\right)\tau_x^2 L^3}.
\end{equation}
Alternatively, we can rearrange Eq.~\ref{eq:emittance-limit-geometric-effects} to become a lower limit on the matched beta function:
\begin{equation}
    \label{eq:beta-function-limit-geometric-effects}
    \beta_0^{\min} \gtrsim \sqrt{\sqrt{18}\left(1+\frac{L}{l}\right)\left(\frac{l}{L}+1\right)\frac{\varepsilon _n\tau_x^2 L^3}{\gamma}}.
\end{equation}
As an example, for the working point in Table~\ref{tab:working-point}, the emittance upper limit is \SI{50}{\milli\meter\milli\radian} for a matched beta function of \SI{15}{mm}, whereas the matched beta function lower limit is \SI{7}{mm} for an emittance of \SI{10}{\milli\meter\milli\radian}.

\begin{figure*}[t]
	\centering
    \includegraphics[width=\linewidth]{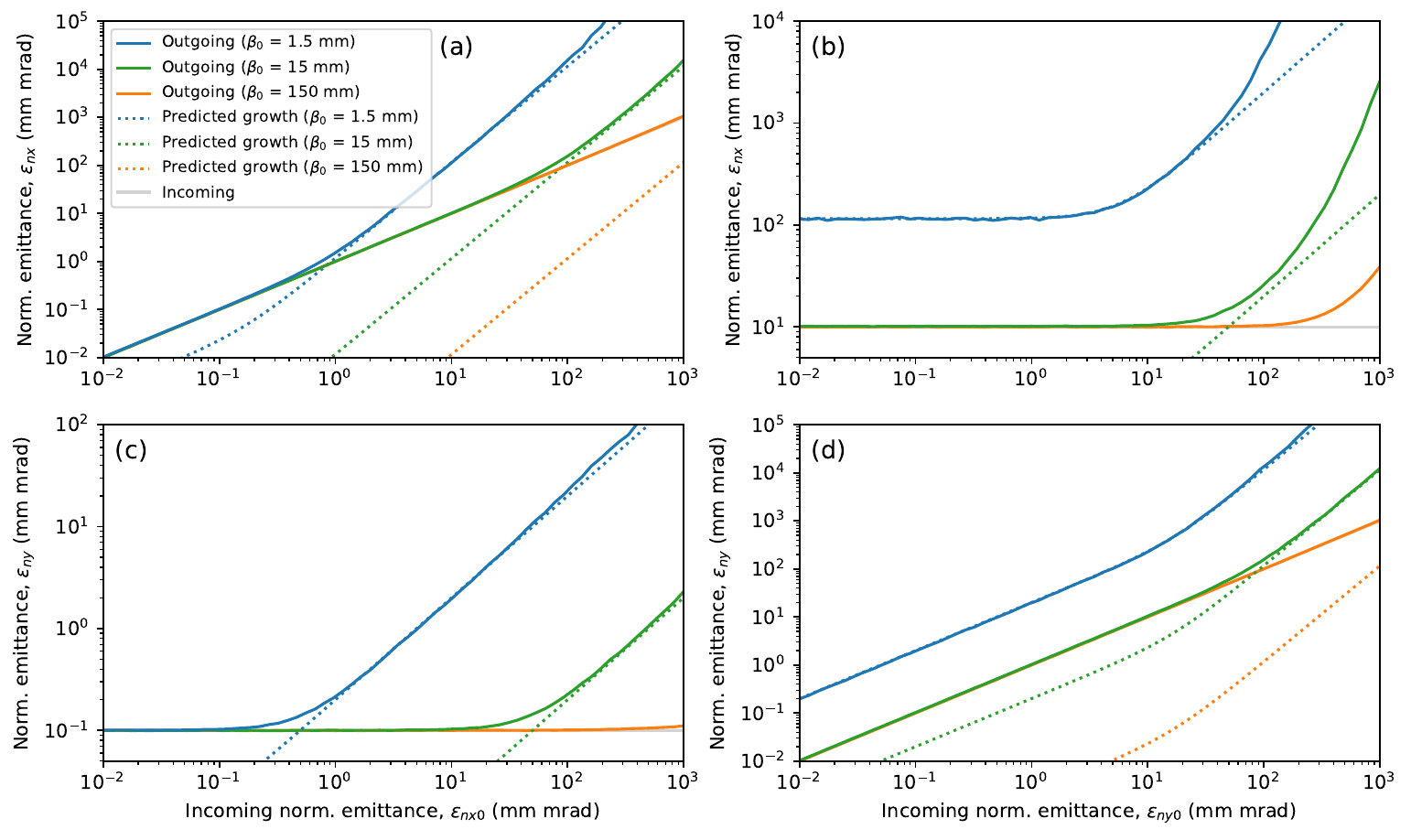}
    \caption{Emittance growth in the horizontal (a,b) and vertical plane (c,d) from geometric aberrations in the nonlinear plasma lenses, shown for scans of the incoming beam emittance in the horizontal (a,c) and vertical plane (b,d). Simulations are shown with solid lines and analytic predictions (Eqs.~\ref{eq:geometric-aberration-x} and \ref{eq:geometric-aberration-y}) are shown with dotted lines. These simulations are performed at \SI{10}{GeV} and zero energy spread. Generally, higher initial beta functions and lower initial emittances lead to less emittance growth, as this minimizes the beam size in the nonlinear plasma-lens field.}
    \label{fig:emittance-scan}
\end{figure*}

Figure \ref{fig:emittance-scan} shows the emittance growth from geometric aberrations in the nonlinear plasma lenses, using simulations scanning the incoming horizontal and vertical emittances at three different matched beta functions (1/10$\times$, 1$\times$ and 10$\times$ compared to that in Table~\ref{tab:working-point}). The analytic prediction matches well, with the exception of the horizontal emittance growth due to high incoming vertical emittances [Fig.~\ref{fig:emittance-scan}(b)], where higher orders of nonlinearity [i.e., $\mathcal{O}(\tau_x^3)$] start to play an important role.

The effect of geometric aberrations in the quadrupole-based staging lattice is not shown here, but has a similar emittance growth to that in the plasma-lens-based lattice; remaining roughly similar in $x$ and approximately a factor 3 higher in $y$ due to the initial defocusing and comparatively larger beam size in this plane. A small vertical emittance growth caused by this increased geometric aberration, not seen in the plasma-lens-based lattice, can be seen in both Fig.~\ref{fig:beam-quality-preservation-quads}(a) and in Fig.~\ref{fig:chromaticity}(b).


\subsection{Misalignment tolerance}
\label{sec:tolerances}

The transverse misalignment of beamline elements can lead to emittance growth, both directly, via the sampling of incorrect nonlinear fields, and indirectly, by an induced centroid offset and/or angle at the end of the achromatic lattice ($\Delta x$ and $\Delta x'$, respectively) that lead to emittance growth in subsequent plasma stages. The latter effect is typically dominant, and can be quantified through the \textit{induced action}~\cite{Lindstrom2016c}
\begin{equation}
    \label{eq:action}
    J_x = \frac{\gamma}{2}\left(\frac{\Delta x^2}{\beta_0} + \beta_0 {\Delta x'}^2\right),
\end{equation}
assuming a beam matched to $\beta_0$ and zero alpha function.

\begin{figure*}[t]
	\centering
    \includegraphics[width=\linewidth]{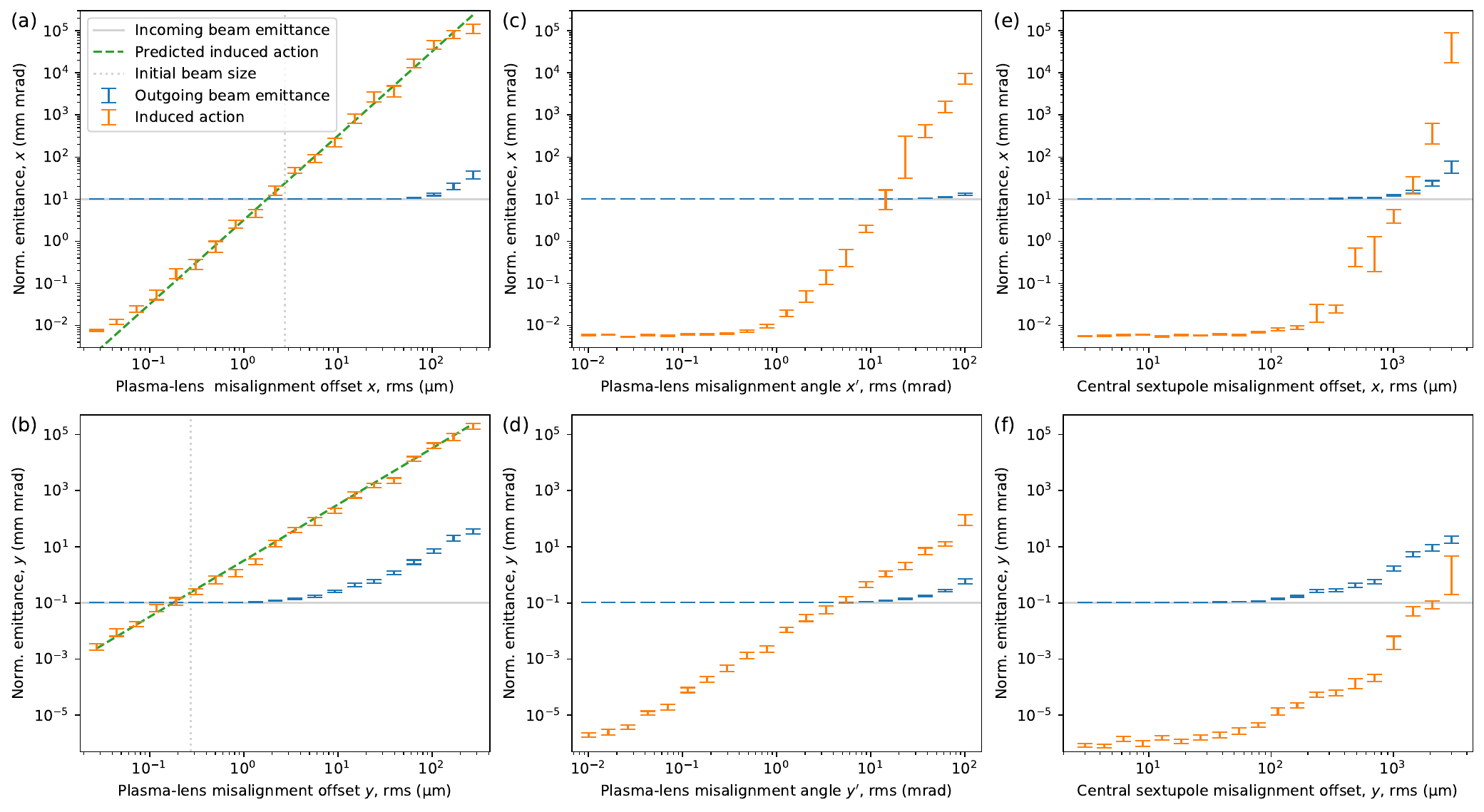}
    \caption{Emittance growth from misalignment of the plasma lenses and central sextupole, simulated at \SI{10}{GeV}, indicating the misalignment tolerances in both $x$ (top) and $y$ (bottom) to plasma-lens offsets (a,b) and angles (c,d) as well as sextupole offsets (e,f). Both the incoming emittance (solid gray line), outgoing emittance (blue error bars) and the induced action (orange error bars) are shown, the latter indicating the expected emittance growth in the subsequent plasma stage. An analytical model for the induced action is shown (dashed green line; Eq.~\ref{eq:prediced-action}); it is in close agreement with the simulations. For plasma lens offsets, the tolerance is set by the induced action and corresponds approximately to the initial beam size (dotted gray line; Eq.~\ref{eq:misalignment-tolerance}). Each error bar indicates the average and standard error from 32 simulations with a different random seed. Note the different misalignment ranges for the plasma lens and sextupoles, indicating the comparatively much higher tolerance to sextupole misalignments.}
    \label{fig:tolerances}
\end{figure*}

Figure~\ref{fig:tolerances} shows the emittance growth for misaligned plasma lenses and for a misaligned central sextupole. Misaligned dipoles are not considered. The figure indicates that the most critical tolerance is that of the plasma-lens offset, particularly here in the vertical plane as this is the plane with the lowest initial emittance. The dominant source of emittance growth is the induced action, and not the direct increase of the emittance. In short, the tolerance in each plane is similar to the initial beam size in that plane; here about \SI{2}{\micro\m} and \SI{0.2}{\micro\m} rms in the $x$ and $y$ planes, respectively. This follows from the expression for the action due to plasma-lens misalignments (derived in Appendix~\ref{app:induced-action})
\begin{equation}
    \label{eq:prediced-action}
    \langle J_x \rangle \approx \frac{\gamma}{\beta_0}\sigma_{\Delta x}^2\left(1+\frac{L}{l}\right)^2,
\end{equation}
where $\sigma_{\Delta x}$ is the rms of the lens offset, assumed to be random, normally distributed and uncorrelated between the two lenses. This implies a misalignment tolerance of
\begin{equation}
    \label{eq:misalignment-tolerance}
    \sigma_{\Delta x}^{\mathrm{max}} \lesssim \sigma_{x0}\left(1+\frac{L}{l}\right)^{-1},
\end{equation}
where $\sigma_{x0}$ is the initial horizontal beam size, assuming we wish to keep the induced action smaller than the initial emittance. For our working point, the misalignment tolerance is 65\% of the beam size. Equation~\ref{eq:misalignment-tolerance} also applies to the $y$ plane, substituting in the vertical beam size, $\sigma_{y0}$.

The tolerance to sextupole misalignment is found to be much less strict compared to that of the plasma lenses. Vertical misalignment causes the most emittance growth, but only starting at around \SI{100}{\micro\m} rms offset. Horizontal misalignment only affects the beam around \SI{1}{\milli\m} rms.

Two further points are worth noting. First, the critical tolerance is given only by the beam size and is not dependent on lens parameters, implying that the tolerance would be similar regardless of what optics are used: plasma lenses are neither better nor worse compared to quadrupole optics. Second, since the two lenses affect the final offset and angle differently (see Appendix~\ref{app:induced-action} for details), this means that offsets of the two plasma lenses can be used in place of correctors. With two degrees of freedom in both planes (i.e., moving both lenses in $x$ and $y$), no additional correctors are in principle necessary.

\subsection{Emittance growth from plasma-specific effects}
\label{sec:emittance-growth-plasma}
Two effects can lead to emittance growth in the plasma lenses: Coulomb scattering and plasma wakefields. These are both discussed below, but only briefly, as their relevance and influence depends strongly on the specific plasma-lens implementation used (see Sec.~\ref{sec:nonlinear-plasma-lenses}).

\subsubsection{Coulomb scattering}

If there are multiple Coulomb scattering events, the expected emittance growth is given by~\cite{Montague1985,Zhao2020,Zhao2022}
\begin{equation}
    \label{eq:scattering}
    \Delta\varepsilon _n \approx 0.83 r_e^2 \left(\frac{\beta_{\mathrm{lens}} L_\mathrm{lens}}{\gamma} \right)n_\mathrm{lens} Z (Z+1) \ln\left(\frac{287}{\sqrt{Z}}\right), 
\end{equation}
where $r_e$ is the classical electron radius, $Z$ is the atomic number of the gas species, $\beta_{\mathrm{lens}}$ is the beta function inside the plasma lens, and $L_{\mathrm{lens}}$ and $n_{\mathrm{lens}}$ are its length and atomic gas density. However, the plasma lenses are typically too short for multiple Coulomb scattering events to occur. In this case, a fraction of the particles receive a kick, while leaving the rest unaffected. Regardless, it is instructive to consider Eq.~\ref{eq:scattering}, as it can be thought of as an upper bound. 

Looking at the example in Table~\ref{tab:working-point}, with an energy of \SI{10}{GeV}, a beta function in the lens of around \SI{70}{m}, a plasma lens of length \SI{50}{mm}, filled with nitrogen ($Z=7$) at a gas density of \SI{1e17}{\per\cm\cubed}, the predicted emittance growth per lens is about \SI{0.03}{\milli\meter\milli\radian}. Due to the random nature of the scattering, the emittance growth is added in quadrature; in the example, this implies increasing a $10\times\SI{0.1}{\milli\meter\milli\radian}$ beam to $<10.00004\times\SI{0.104}{\milli\meter\milli\radian}$, which is negligible.

\subsubsection{Distortion from wakefield focusing}
In an active plasma lens, if the beam density is sufficiently high, it can drive a plasma wake whose transverse focusing forces compete with the lens field, causing distortion and possibly emittance growth~\cite{Pompili2018,Lindstrom2018b}. However, this does not apply to passive plasma lenses, as these already rely on a plasma wakefield (driven by another laser or particle beam) to provide the focusing.

The maximum focusing strength within the plasma wakefield, in units of magnetic field gradient, can be expressed as~\cite{Lindstrom2018b}
\begin{equation}
    \label{eq:wakefield-distortion}
    g_\mathrm{wake}^\mathrm{max} \approx -\frac{c \mu_0 Q k_p^2 \sigma_z}{2\pi\sigma_x\sigma_y(1+k_p^2\sigma_x\sigma_y/2)(1+\sqrt{2\pi}k_p^2\sigma_z^2)}, 
\end{equation}
where $Q$ is the beam charge, $\sigma_z$ is the bunch length, $\sigma_x$ and $\sigma_y$ are the transverse beam sizes in the lens, and $n_0$ is the plasma density. Equation~\ref{eq:wakefield-distortion} assumes linear plasma wakefields---in the blowout regime, the gradient is capped at $-e c \mu_0 n_0/2$. Note also that the focusing gradient (Eq.~\ref{eq:wakefield-distortion}) is not the average but the peak (typically occurring at the center of the beam tail), and therefore most of the beam particles will see a smaller gradient. In practice, therefore, one must perform a particle-in-cell simulation to understand the effect on the emittance.

Using the parameters in Table~\ref{tab:working-point}, where the transverse beam size reaches $370\times\SI{19.5}{\micro\meter}$ in the lens [see Fig.~\ref{fig:phase-space-evolution}(c)], and assuming a density of \SI{e17}{\per\cm\cubed}, Eq.~\ref{eq:wakefield-distortion} estimates a peak focusing gradient of \SI{276}{T/m} or about 28\% of the APL field. This is small but not negligible. 

In conclusion, the wakefield focusing effect in APLs and how it interacts with the achromatic lattice must be carefully considered, and is therefore the subject of an upcoming study. Mitigation methods include operating with higher currents in a shorter APL, to reduce the relative effect, or simply switching to a passive plasma lens.


\section{Synchrotron radiation and energy scaling}
\label{sec:energy-scaling-and-sr}

Synchrotron radiation, both coherent and incoherent, will affect any high-energy lattice that contains strong dipole magnets. The coherence length of the synchrotron radiation can at low energy be comparable to the bunch length, resulting in coherent emission; coherent synchrotron radiation (CSR)~\cite{Saldin1997}. CSR is a multi-particle effect and will therefore place limits on the charge and bunch length at low energies, as discussed in Sec.~\ref{sec:csr-limits} below. Conversely, at high energies, incoherent synchrotron radiation (ISR) will dominate. 

Another important consideration, also affected by synchrotron radiation, is how the lattice (i.e., the lengths and strengths of its elements) should be scaled with energy. Section~\ref{sec:energy-scaling} presents a basic scaling solution, which works for energies below \SI{\sim 50}{GeV}, Sec.~\ref{sec:effect-of-csr-and-isr} simulates the effects of CSR and ISR across a wide range of energy, and Sec.~\ref{sec:improved-energy-scaling} presents an improved solution with decreasing dipole magnetic fields in order to suppress emittance growth from ISR for energies above this energy threshold.

\subsection{CSR limit on the beam's peak current}
\label{sec:csr-limits}

\begin{figure}[b]
	\centering
    \includegraphics[width=\linewidth]{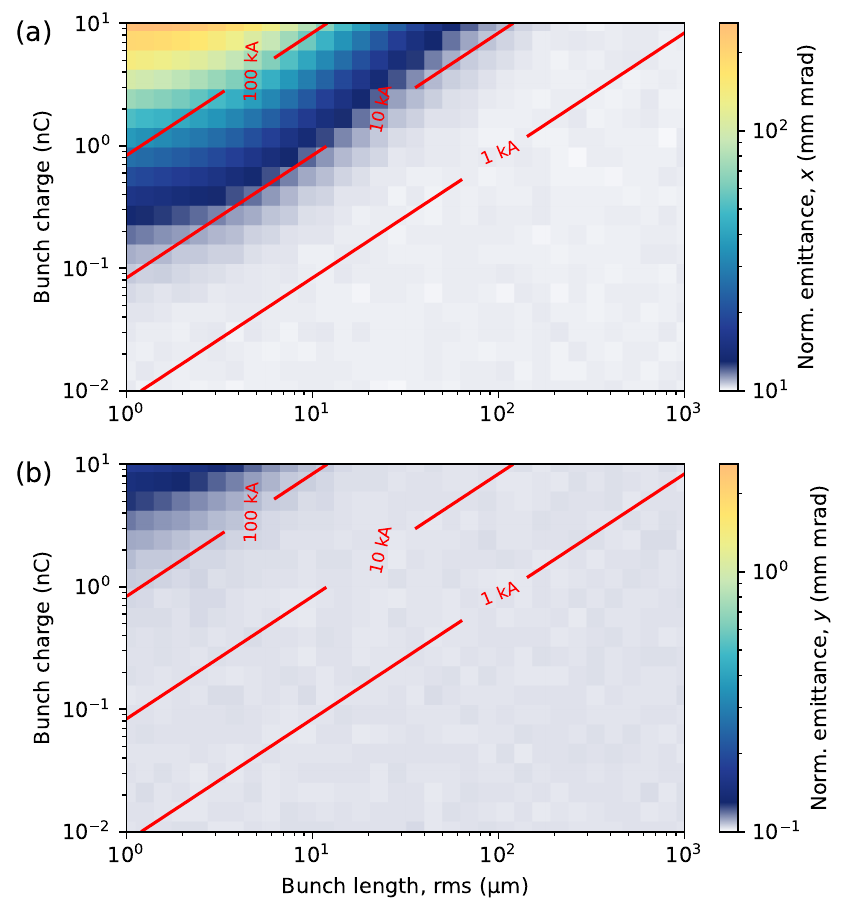}
    \caption{Simulated effect of coherent synchrotron radiation on emittance in the horizontal (a) and vertical plane (b), for different combinations of bunch length and charge. The resulting emittance after traversing a single achromatic lattice at \SI{10}{GeV} (white-to-rainbow color map) only grows for high charges and short bunches. Corresponding peak currents are indicated (red contour lines), showing that emittance is preserved up to approximately \SI{10}{kA} at this particular energy and main dipole field strength (i.e., \SI{10}{GeV} and \SI{1}{T}, respectively). The effect is most prominent in the bending plane (i.e., $x$), but starts to affect the non-bending plane (i.e., $y$) when the effect is sufficiently large. }
    \label{fig:csr-limits}
\end{figure}

Figure~\ref{fig:csr-limits} shows the effect of CSR on the emittance in both transverse planes as we vary the bunch length and charge. \textsc{ImpactX}, used for these simulations, uses a 1D ultrarelativistic steady-state wakefield model~\cite{Saldin1997} for CSR. For the particular beam parameters (i.e., those in Table~\ref{tab:working-point}), we find there is an approximate limit on the peak current of about \SI{10}{kA}. The peak current in the standard example is \SI{2}{kA}, which was specifically chosen to avoid emittance growth from CSR. However, the 2D limit in bunch charge versus bunch length is not uniquely determined by their ratio (i.e., peak current); the relationship is somewhat more complex and should therefore be determined by simulation for any given setup.

\subsection{Basic energy-scaling lattice solution}
\label{sec:energy-scaling}

It is possible to imagine several ways of scaling the beamline-element lengths and strengths to work with higher energy beams. One basic solution that stands out as particularly simple and practical is summarized in Table~\ref{tab:energy-scaling-nominal}. It applies equally to the plasma-lens-based and the alternative quadrupole-based lattices, and is justified below.

For the focusing elements, assuming that the plasma lenses are operated at the highest available magnetic-field gradient $g_0$ at any energy, the focusing strength will scale inversely with energy: $k_\mathrm{lens} = g_0 q c / \mathcal{E} \sim 1/\mathcal{E}$. This means that the focal length scales as $f = 1/(k_\mathrm{lens} L_\mathrm{lens}) \sim \mathcal{E}/L_\mathrm{lens}$. A solution that minimizes the overall length of the lattice, and results in a similar optical solution, is to scale both the length and focal length of the plasma lens identically, i.e.~as $f \sim L_\mathrm{lens} \sim \sqrt{\mathcal{E}}$. More generally, all element lengths should be scaled with $\sqrt{\mathcal{E}}$. If the initial beta function also scales this way, $\beta_0 \sim \sqrt{\mathcal{E}}$, as it does naturally when matched in a plasma accelerator, then the evolution of the beta function is identical at any energy (but scaled with $\sqrt{\mathcal{E}}$ everywhere).

\begin{table}[b]
    \centering
    \begin{tabular}{lccc}
         & & \textit{Basic} & \textit{Improved} \\
        \textit{Parameter (unit)} & \textit{Symbol} & \textit{scaling} & \textit{scaling}\\
        \hline
        Element lengths (m) & $L$ & $\sqrt{\mathcal{E}}$ & $\sqrt{\mathcal{E}}$\\
        Beta function (m) & $\beta$ & $\sqrt{\mathcal{E}} $ & $\sqrt{\mathcal{E}} $\\
        Dipole fields (T) & $B$ & $1$ & $\mathcal{E}^{-3/5}$\\
        Plasma-lens field (T/m) & $g_0$ & $1$ & $1$\\
        Plasma-lens nonlinearity (1/m) & $\tau_x$ & $1$ & $\mathcal{E}^{3/5}$\\
        Sextupole field (T/m$^2$) & $m_{\mathrm{sext}}$ & $1$ & $\mathcal{E}^{3/5}$ \\
        Horizontal disp., any order (m) & $D^{(n)}_x$ & $1$ & $\mathcal{E}^{-3/5}$\\
        Angular dispersion, any order & $D^{(n)}_{x'}$ & $1/\sqrt{\mathcal{E}}$ & $\mathcal{E}^{-11/10}$\\
        Longitudinal dispersion (m) & $R_{56}$ & $1/\sqrt{\mathcal{E}}$ & $\mathcal{E}^{-17/10}$\\
        Total bending angle, lattice (rad) & $\theta$ & $1/\sqrt{\mathcal{E}}$ & $\mathcal{E}^{-11/10}$ \\
        \hline
    \end{tabular}
    \caption{Energy ($\mathcal{E}$) scaling of key parameters for the basic and improved energy-scaling solutions.}
    \label{tab:energy-scaling-nominal}
\end{table}

If the dipoles, whose length should scale as $L_0 \sim L_\mathrm{chic} \sim \sqrt{\mathcal{E}}$, maintain the same $B$-field at all energies, this will result in a dispersion in the plasma lens that is independent of the energy ($D_x \sim B L^2/\mathcal{E}\sim\mathrm{const}$; see Eq.~\ref{eq:dispersion-in-lens}). This implies that the nonlinearity in the plasma lens is also constant ($\tau_x = 1/D_{x,\mathrm{lens}}\sim\mathrm{const}$; Eq.~\ref{eq:nonlinearity-matching})---this is important for implementation purposes, as all lenses operate identically and only vary in length. Looking at the bending angle of each element, and therefore also the whole lattice, this will scale as $\theta = BLec/\mathcal{E} \sim 1/\sqrt{\mathcal{E}}$. Further, this also implies that the angular dispersion $D_{x'}$ (which scales with the bending angle) will scale as $1/\sqrt{\mathcal{E}}$. 

The longitudinal dispersion, $R_{56} \sim B^2 L^3 / \mathcal{E}^2$ (see Eq.~\ref{eq:R56-main-dipole}), will scale as $1/\sqrt{\mathcal{E}}$. This has implications for the employment of the longitudinal self-correction mechanism~\cite{Lindstrom2021b}, suggesting that the self-correction process needs to have converged while the accelerated beam is at low energy (i.e., the first stages), before the longitudinal phase space is effectively ``locked in'' at high energy.

The central sextupole should also scale in length as $L_\mathrm{sext} \sim \sqrt{\mathcal{E}}$ while maintaining the same magnetic gradient $m_\mathrm{sext}$. Following the above scaling, the magnitude of the second-order dispersion is independent of energy, just like the first-order dispersion. This works because the integrated sextupole strength $k_{2,\mathrm{sext}} L_{\mathrm{sext}} = m_\mathrm{sext} L_{\mathrm{sext}} e c /\mathcal{E} \sim D^{(2)}_{x',\mathrm{sext}}/D_{x,\mathrm{sext}}$ scales as $1/\sqrt{\mathcal{E}}$, which is consistent with a constant $m_\mathrm{sext}$.

Finally, the driver separation (see Sec.~\ref{sec:driver-separation}) scales differently for laser and beam drivers. A beam driver, whose energy $\mathcal{E}_\mathrm{d}$ is constant even as the nominal lattice energy and dipole length increases, will be deflected at an increasing angle, scaling as $\theta_\mathrm{d} = BLec/\mathcal{E}_\mathrm{d} \sim \sqrt{\mathcal{E}}$ until the point where the driver exits the dipole (at which point the angle no longer increases). This also increases the transverse separation of the driver and the plasma lens (see Eq.~\ref{eq:beam-driver-separation}), scaling as $(\mathcal{E}/\mathcal{E}_\mathrm{d}-1)$, again until the driver exits the dipole. A laser driver, however, will be separated from the lens by the same distance regardless of energy (since $D_{x,\mathrm{lens}}$ stays constant), but at a distance increasingly further away from the stage (scaling as $\sqrt{\mathcal{E}}$); this means that the allowable laser divergence decreases with energy, scaling as $1/\sqrt{\mathcal{E}}$ (same as the angular dispersion)---an unfavorable scaling at high energy.

\subsection{Effect of CSR and ISR at different energies}
\label{sec:effect-of-csr-and-isr}

\begin{figure*}[t]
	\centering
    \includegraphics[width=\linewidth]{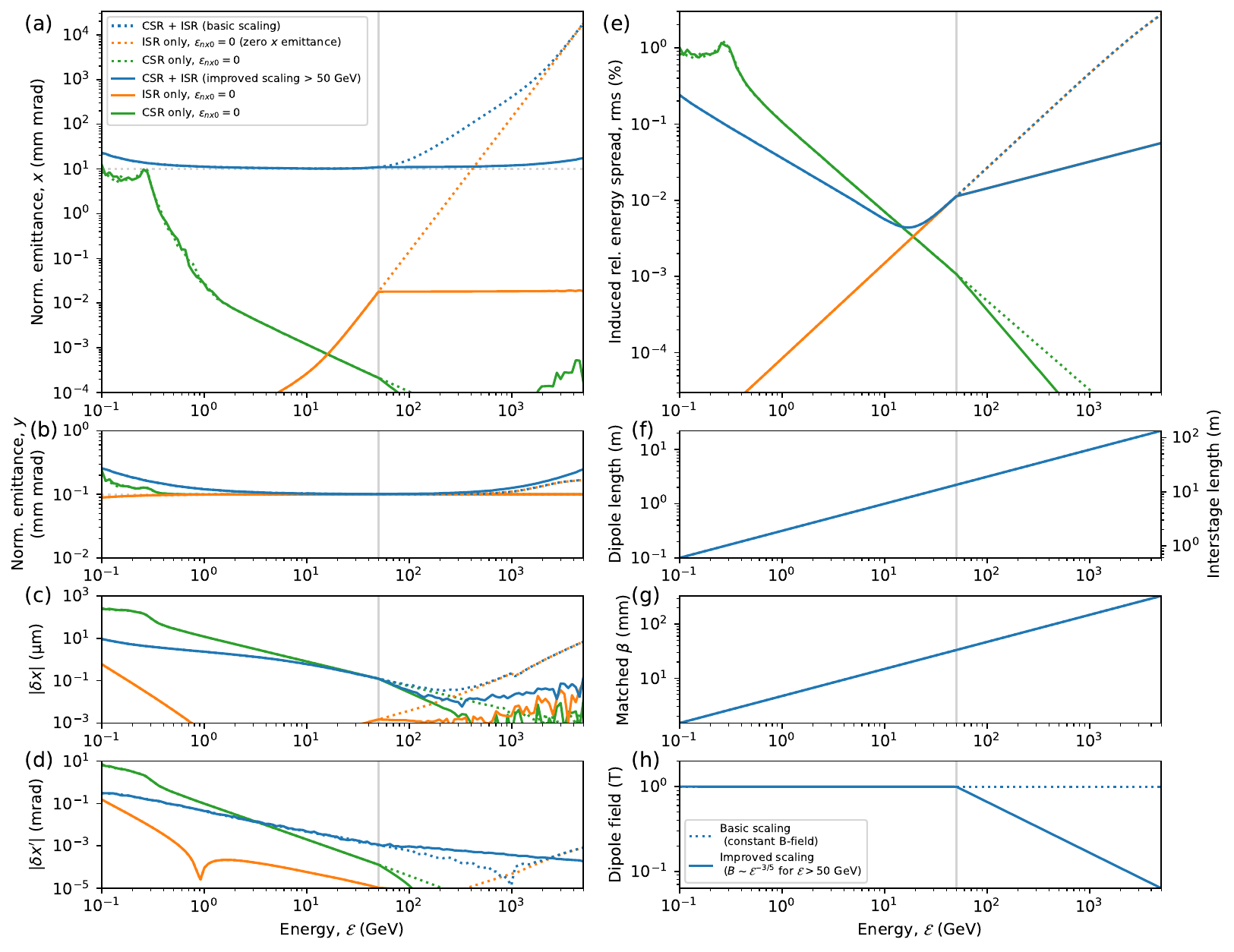}
    \caption{Energy scaling, showing the simulated effect of CSR and ISR on various beam parameters (a--e) based on scaled lattice parameters (f--h). The blue line shows a realistic beam with non-zero emittance (10 mm mrad) and both CSR and ISR enabled. The green and orange lines represent a beam with zero initial horizontal emittance in order to observe the contribution from CSR and ISR separately. The solid lines represent a basic energy scaling with constant B-field, whereas the dotted lines represent an improved scaling at high energy (above 50 GeV) that ramps down the B-field with energy to ensure constant horizontal emittance growth. $|\delta x|$ and $|\delta x'|$ refer to the induced centroid offset and angle at the end of the lattice, relative to a reference particle (also affected by ISR when enabled). The initial energy spread is set to zero in all simulations.}
    \label{fig:energy-scaling}
\end{figure*}

Figure~\ref{fig:energy-scaling} shows a simulation of the effect on key beam parameters of scaling the energy. The energy ranges from \SI{100}{MeV} to \SI{5}{TeV}, with the maximum energy chosen for its relevance to a \SI{10}{TeV} center-of-mass wakefield-based collider~\cite{Gessner2025}. Simulations are performed with the plasma-lens-based working point (Table~\ref{tab:working-point}), scaling parameters with energy as per Table~\ref{tab:energy-scaling-nominal}; both the basic scaling and an improved scaling is used (discussed in Sec.~\ref{sec:improved-energy-scaling} below). Additionally, simulations are performed with zero horizontal emittance in order to showcase separately the emittance growth from CSR and ISR.

CSR affects the beam quality mostly at low energies, as discussed in Sec.~\ref{sec:csr-limits}. Emittance growth from CSR is non-trivial to model analytically and has a complex energy scaling, but generally decreases with energy. At high energies the emittance growth in the horizontal plane can be seen to scale as $\mathcal{E}^{-3/2}$. Ultimately, this means that at sufficiently high energies (here above \SI{1}{GeV}), CSR can be completely neglected. At sufficiently low energy (here below \SI{300}{MeV}), however, the growth in emittance and energy spread is disruptive. Being a collective effect, CSR can however be suppressed by reducing the peak current (as shown in Fig.~\ref{fig:csr-limits}).

ISR, on the other hand, grows with energy and is therefore negligible at lower energies (here below \SI{50}{GeV}). Since ISR is a single-particle effect it is simpler to model, but still somewhat non-trivial as the emittance growth can depend on the beam size evolution, which again depends on the emittance growth. At sufficiently high energies, however, the emittance growth in the horizontal (bending) plane from ISR alone will scale as~\cite{Raubenheimer1993}
\begin{equation}
    \label{eq:isr-emittance-scaling}
    \Delta \varepsilon _{nx}^\mathrm{ISR} \sim \mathcal{E} L^4 B^5,
\end{equation}
assuming that the beta function scales with the lattice length. Using the basic length scaling of $L\sim\sqrt{\mathcal{E}}$ and a constant B-field, we get $\Delta \varepsilon _{nx} \sim \mathcal{E}^3$. This rapid growth with energy is problematic at high energy and must therefore be mitigated (see Sec.~\ref{sec:improved-energy-scaling} below). Note that strong-field quantum-electrodynamics effects, potentially relevant at high energies, are included in these \textsc{ImpactX} simulations up to third order in the quantum nonlinearity parameter $\chi$~\cite{Gonoskov2022}, but no significant effect is observed.

Finally, both CSR and ISR can introduce centroid offsets ($|\delta x|$) and kicks ($|\delta x'|$). This can cause transverse oscillations in the subsequent plasma stage, leading to emittance growth~\cite{Lindstrom2016c,Thevenet2019} and seeding transverse instabilities~\cite{Lebedev2017,Finnerud2025}. However, we can safely ignore these effects since it is possible to correct them by transversely offsetting the two plasma lenses, using them as correctors (see Sec.~\ref{sec:tolerances} and Appendix~\ref{app:induced-action}).

\subsection{Improved energy scaling for ISR suppression at high energy}
\label{sec:improved-energy-scaling}

If we instead wish to maintain a constant emittance growth from ISR for increasing energy, we find from Eq.~\ref{eq:isr-emittance-scaling} that we need to ramp down the dipole magnetic fields as $B\sim \mathcal{E}^{-3/5}$. This is referred to here as the \textit{improved scaling}, and is only applied above a certain energy threshold [see Fig.~\ref{fig:energy-scaling}(h)]. The resulting energy scaling of each parameter is shown in Table~\ref{tab:energy-scaling-nominal}.

Since ISR emittance growth can depend on the initial emittance, there exists a transition regime of intermediate energies (here from \SI{50}{GeV} to \SI{2}{TeV}) where it is not straightforward to predict the emittance growth analytically. Instead the exact energy threshold should be determined by simulation---for our example parameters this threshold is \SI{50}{GeV}.

The improved scaling comes with the added complexity of increasing the plasma-lens nonlinearity with energy ($\tau_x \sim \mathcal{E}^{3/5}$). This has two downsides: firstly, implementing such strongly nonlinear plasma lenses may be challenging to achieve in practice, at least in APLs but perhaps less so in PPLs; secondly, and more importantly, at sufficiently high energy the nonlinearity is so strong that geometric effects lead to emittance growth [given by Eq.~\ref{eq:geometric-aberration-x} and seen in Fig.~\ref{fig:energy-scaling}(a)]. Nevertheless, in the example, this emittance growth is not very significant until energies around 1--\SI{5}{TeV}. 

The reduced magnetic field also affects driver separation. It exacerbates further the limit on the maximum divergence of a laser driver, now scaling as $\mathcal{E}^{-11/10}$ (like the angular dispersion), prompting the need for a plasma mirror~\cite{Thaury2007,Steinke2016,Zingale2021} close to the stage instead of letting the laser pass the plasma lens. Beam drivers, on the other hand, continue to increase their separation, albeit more slowly, scaling as $\mathcal{E}^{-3/5}(\mathcal{E}/\mathcal{E}_\mathrm{d}-1)\approx\mathcal{E}^{2/5}$.

Ultimately, this improved scaling increases the energy range from the original two orders of magnitude (i.e., 0.5--\SI{50}{GeV}) to about four orders of magnitude (i.e., 0.5--\SI{5000}{GeV}).


\section{Conclusions}
\label{sec:conclusions}

In this paper, we have introduced and explored an achromatic lattice based on nonlinear plasma lenses, which can transport beams with high divergence and large energy spreads. Being both simpler and shorter than an equivalent quadrupole-based lattice, it also outperforms in terms of transportable energy bandwidth; up to several percent rms. The lattice has a tunable $R_{56}$ to allow for longitudinal self-correction in a multistage plasma accelerator, providing intrinsic energy stability and reduced energy spread, and can be scaled in energy over 4 orders of magnitude, from \SI{0.5}{GeV} or lower to \SI{5}{TeV} or higher. 

In short, this achromatic lattice promises to solve the staging problem in plasma acceleration.

\acknowledgements
The authors would like to thank J. Bj{\"o}rklund Svensson, L. Verra, A. Knetsch, S. Boogert, B. Foster and J. Osterhoff for valuable discussions and inputs. This work is funded by the European Research Council (ERC Grant No.~101116161) and the Research Council of Norway (NFR Grant No.~313770).
A.H.~and C.E.M.~are supported by the CAMPA collaboration, a project of the U.S.~Department of Energy, Office of Science, Office of Advanced Scientific Computing Research and Office of High Energy Physics, Scientific Discovery through Advanced Computing (SciDAC) program.
The authors acknowledge all \textsc{ABEL} and \textsc{ImpactX} contributors.

\appendix

\section{Derivation of the magnetic field profile in a nonlinear plasma lens}
\label{app:magnetic-field-profile}
Starting from the focusing strengths in Eqs.~\ref{eq:force-gradient-x} and ~\ref{eq:force-gradient-y}, and assuming an active plasma lens (where $\mathbf{B}\neq0$ but $\mathbf{E}=0$), we find that the magnetic field gradients are
\begin{align}
    \label{eq:magnetic-gradient-y}
    \frac{\partial B_y}{\partial x} &= \frac{k_0 \mathcal{E}_0}{q c} \left(1 + \tau_x x\right), \\
    \label{eq:magnetic-gradient-x}
    \frac{\partial B_x}{\partial y} &= -\frac{k_0 \mathcal{E}_0}{q c} \left(1 + \tau_x x\right).
\end{align}
Both $B_x$ and $B_y$ can be functions of position in $x$ and $y$, which means that the integral of the partial derivatives (e.g., in $x$) will have constants that are functions of the non-integrated variable [e.g., $c(y)$]. Let us first integrate Eq.~\ref{eq:magnetic-gradient-y} to find
\begin{equation}
    \label{eq:magnetic-field-y}
    B_y(x,y) = g_0 \left(x +\tau_x \frac{x^2}{2}\right) + f_y(y),
\end{equation}
where the $g_0 = k_0\mathcal{E}_0/qc$ is a nominal magnetic field gradient and $f_y(y)$ is an unknown function of only $y$ (but not of $x$). Next, we integrate Eq.~\ref{eq:magnetic-gradient-x} to find 
\begin{equation}
    \label{eq:magnetic-field-x}
    B_x(x,y) = -g_0 \left(y + \tau_x x y\right) + f_x(x),
\end{equation}
where similarly $f_x(x)$ is an unknown function of only $x$ (but not of $y$).

To determine the unknown functions $f_x(x)$ and $f_y(y)$, we make use of Gauss's law for magnetism:
\begin{eqnarray}
    \nabla \cdot \mathbf{B} &=& 0 \\
    \frac{\partial B_x}{\partial x} + \frac{\partial B_y}{\partial y} + \frac{\partial B_z}{\partial z} &=&  0 \\
    \label{eq:gauss-law-magnetic-field-line3}
    -g_0 \tau_x y + \frac{\partial f_x(x)}{\partial x} + \frac{\partial f_y(y)}{\partial y} &=&  0,
\end{eqnarray}
where we have used the assumption that $B_z = 0$ everywhere. The first and third terms of Eq.~\ref{eq:gauss-law-magnetic-field-line3} are only functions of $y$, whereas the second term is only a function of $x$. However, their sum must always equal zero, regardless of the value of $x$ and $y$. This implies that
\begin{eqnarray}
    \frac{\partial f_y(y)}{\partial y} &=& g_0 \tau_x y, \\
    \frac{\partial f_x(x)}{\partial x} &=&  0,
\end{eqnarray}
which can be integrated to give
\begin{eqnarray}
    f_y(y) &=& \frac{g_0 \tau_x y^2}{2} + B_{y0}, \\
    f_x(x) &=& B_{x0},
\end{eqnarray}
where $B_{x0}$ and $B_{y0}$ are constants that can, in the next step, be interpreted as horizontal and vertical dipole fields, respectively. Finally, inserting $f_x(x)$ and $f_y(y)$ back into Eqs.~\ref{eq:magnetic-field-x} and \ref{eq:magnetic-field-y}, respectively, we get
\begin{eqnarray}
    B_x &=& -g_0 \left(y + \tau_x x y\right) + B_{x0}, \\
    B_y &=& g_0 \left(x +\tau_x \frac{x^2+y^2}{2}\right) + B_{y0}.
\end{eqnarray}

A very similar prescription can be followed to show the electric fields in a nonlinear passive plasma lens (where instead $\mathbf{E}\neq0$ but $\mathbf{B}=0$), resulting in Eqs.~\ref{eq:nonlinear-electric-field-x} and \ref{eq:nonlinear-electric-field-y}.

\section{Derivation of the transverse and longitudinal dispersions induced in the main dipole}
\label{app:main-dipole-dispersions}

We start by defining the transverse dispersion of order $n$, denoted $D^{(n)}_x$, as the coefficient of the Taylor series in relative energy offset $\delta$ for the transverse offset
\begin{equation}
    \label{eq:dispersion-definition}
    x(\delta) = x(0) + D_x \delta + D^{(2)}_x \delta^2 + ...
\end{equation}
where $x(0)$ is the initial offset and $D_x\equiv D^{(1)}_x$. Alternatively, in derivative form, 
\begin{equation}
    D^{(n)}_x = \frac{1}{n!}\frac{\partial^nx}{\partial \delta^n}. 
\end{equation}

In a dipole with magnetic field $B_0$, a particle of momentum $p$ will undergo circular motion with bending radius $\rho = p/qB_0$. After traversing a longitudinal distance $s$ along the circle, the transverse offset from the initial axis will therefore be
\begin{equation}
    x = \rho\cos\left(\frac{s}{\rho}\right)-\rho = -\frac{s^2}{2\rho} + \mathcal{O}(\rho^{-3}),
\end{equation}
where we have used the small angle approximation. For ultrarelativistic particles, $p\approx\mathcal{E}(1+\delta)/c$, where $\mathcal{E}$ is the nominal energy and $c$ is the speed of light in vacuum. Substituting in $\rho = p/qB_0$, we can expand the above equation as 
\begin{equation}
    \label{eq:dispersion-expanded}
    x \approx -\frac{B_0s^2qc}{2\mathcal{E}(1+\delta)} \approx -\frac{B_0s^2qc}{2\mathcal{E}}\left(1-\delta+\delta^2+\mathcal{O}(\delta^{3})\right).
\end{equation}
Comparing Eqs.~\ref{eq:dispersion-definition} and \ref{eq:dispersion-expanded}, we can read off the first-order transverse dispersion at the end of the main dipole of length $L_0$ (i.e., just before the first plasma lens) as
\begin{equation}
    D_{x,\mathrm{lens}} \approx \frac{B_0L_0^2qc}{2\mathcal{E}},
\end{equation}
and the second-order dispersion as
\begin{equation}
    \label{eq:second-order-dispersion}
    D^{(2)}_{x,\mathrm{lens}} \approx -\frac{B_0L_0^2qc}{2\mathcal{E}} \approx -D_{x,\mathrm{lens}}.
\end{equation}

Next, we consider the longitudinal dispersion, $R_{56}$, which is can be calculated using the integral
\begin{equation}
    R_{56} = \int\frac{D_x(s)}{\rho}\mathrm{d} s.
\end{equation}
Starting again from zero angular and positional dispersion (i.e., $D'_x = D_x = 0$) we get $D_x(s) = s^2/2\rho$. Integrating over the length $L_0$, we find that
\begin{equation}
    R_{56} = \int_0^{L_0}\frac{s^2}{2\rho^2}\mathrm{d}s = \frac{L_0^3}{6\rho^2},
\end{equation}
or equivalently, if substituting in $\rho \approx \mathcal{E}/cqB_0$,
\begin{equation}
    R_{56} \approx \frac{B_0^2 L_0^3 c^2 q^2}{6 \mathcal{E}^2},
\end{equation}
again assuming ultrarelativistic particles.

\section{Derivation of emittance growth from second-order chromaticity}
\label{app:second-order-chromaticity}

In order to estimate the effect of higher-order chromaticity, we simplify the problem by only calculating the emittance growth in the vertical (undispersed) plane and assume that there are no geometric effects (see Appendix~\ref{app:nonlinear-forces} for a separate treatment of these). The same chromatic emittance growth will apply to the horizontal (dispersed) plane, but only at the end of the lattice where dispersions are cancelled.

We start by assuming large energy offsets and a small emittance in the horizontal plane, such that the chromatic terms (i.e., those with $\delta$) dominate the horizontal position $x$ and angle $x'$. The horizontal position in the plasma lens, dominated by the dispersion, will therefore be approximately
\begin{equation}
    x_1 \approx \frac{\delta}{\tau_x} - \frac{\delta^2}{\tau_x} + \mathcal{O}(\delta^3) + \mathcal{O}(x_0) + \mathcal{O}(x'_0),
\end{equation}
where we have expanded the dispersion terms up to second order in $\delta$ (see Appendix~\ref{app:main-dipole-dispersions} and Eq.~\ref{eq:dispersion-definition}), used the relation $D^{(2)}_{x,\mathrm{lens}} \approx -D_{x,\mathrm{lens}}$ between the first and second-order dispersions at the lens (see Eq.~\ref{eq:second-order-dispersion}), and have substituted in the lens nonlinearity $\tau_x$ using $D_{x,\mathrm{lens}} = 1/\tau_x$ (Eq.~\ref{eq:nonlinearity-matching}).

The vertical position in the lens, not affected by dispersion, is simply
\begin{equation}
    y_1 = y_0 + y'_0 L \approx y'_0 L,
\end{equation}
approximated using $y_0 \ll y'_0 L$ (because $\beta_0 \ll L$), where $y_0$ and $y'_0$ are the initial vertical position and angle of the particle and $L$ is the distance to the first lens. The effect of the lens is to change the angle $y' \to y' - y/f$. The nominal focal length of the lens is $f_0 = (L^{-1}+l^{-1})^{-1}$, where $l$ is the distance from the lens to the lattice center. Incorporating the nonlinearity and energy dependence of the lens, we get 
\begin{equation}
    f(x,\delta) = f_0\frac{1+\delta}{1+\tau_x x}.
\end{equation}
Therefore, the angle after the lens becomes
\begin{eqnarray}
    y'_1 &=& y'_0 - \frac{y'_0 L}{f_0}\frac{1+\tau_x x_1}{1+\delta} \\
    &=& y'_0 - y'_0 \left(1+\frac{L}{l}\right)\frac{1+\delta-\delta^2+\mathcal{O}(\delta^3)}{1+\delta} \\
    &=& y'_0 - y'_0 \left(1+\frac{L}{l}\right)\left(1-\delta^2+\mathcal{O}(\delta^3)\right) \\
    &=& -y'_0\frac{L}{l} + y'_0 \left(1+\frac{L}{l}\right)\delta^2 + \mathcal{O}(\delta^3).
\end{eqnarray}
The resulting offset in the second lens, after a drift of length $2l$ (i.e., the distance between the two lenses) is
\begin{eqnarray}
    y_2 &=& y_1 + 2ly'_1 \\
    y_2 &=& y_0 - y'_0 L + 2y'_0L\left(\frac{l}{L}+1\right)\delta^2 + \mathcal{O}(\delta^3).
\end{eqnarray}
Continuing, we calculate the particle's angle after the second lens. This will again depend on the $x$ position in the second lens, again dominated by the first- and second-order dispersion, which we will assume is identical to that in the first lens (this is the point of introducing the chicane dipoles and central sextupoles); i.e., $x_1 \approx x_2$. The vertical angle after the second lens, which is also the angle at the end of the lattice, is therefore
\begin{widetext}
    \begin{eqnarray}
        y'_3 = y'_2 &=& y_1' - \frac{y_2}{f_0}\frac{1+\tau_x x_2}{1+\delta} \\
        &\approx& y_1' - \left(\frac{1}{L}+\frac{1}{l}\right)\left(- y'_0 L + 2y'_0L\left(\frac{l}{L}+1\right)\delta^2\right)\frac{1+\delta-\delta^2}{1+\delta} +\mathcal{O}(\delta^3) \\
        &\approx& -y'_0\frac{L}{l} + y'_0 \left(1+\frac{L}{l}\right)\delta^2 - \left(\frac{1}{L}+\frac{1}{l}\right)\left(-y'_0 L + 2y'_0L\left(\frac{l}{L}+1\right)\delta^2 \right)\left(1-\delta^2\right) +\mathcal{O}(\delta^3) \\
        &\approx& -y'_0\frac{L}{l} + y'_0 \left(1+\frac{L}{l}\right)\delta^2 + \left(1+\frac{L}{l}\right)y'_0\left(1-\delta^2\right) - 2y'_0\left(1+\frac{L}{l}\right)\left(\frac{l}{L}+1\right)\delta^2 +\mathcal{O}(\delta^3) \\
        &\approx& y'_0 - 2y'_0\left(1+\frac{L}{l}\right)\left(\frac{l}{L}+1\right)\delta^2 +\mathcal{O}(\delta^3).
    \end{eqnarray}
\end{widetext}
\begin{widetext}
The corresponding offset at the end of the lattice is
\begin{eqnarray}
    y_3 &=& y_2 + y'_2 L \\
    &=& y_0 - y'_0 L + 2y'_0L\left(\frac{l}{L}+1\right)\delta^2 + y'_0L - 2y'_0L\left(1+\frac{L}{l}\right)\left(\frac{l}{L}+1\right)\delta^2 + \mathcal{O}(\delta^3) \\
    &=& y_0 - 2y'_0L\left(1+\frac{L}{l}\right)\delta^2 + \mathcal{O}(\delta^3). \\
\end{eqnarray}
\end{widetext}
Note that there is no linear term in $\delta$ in either $y_3$ or$y'_3$, which implies that the first-order chromaticity is canceled (as intended). We now calculate the geometric emittance, which is averaged over a bi-Gaussian distribution in $y_3$ and $y'_3$
\begin{widetext}
    \begin{eqnarray}
        \varepsilon _{y} &=& \sqrt{\langle y_3^2\rangle\langle {y'_3}^2\rangle - \langle y_3 {y'_3}\rangle^2} \\
        &=& \sqrt{\langle (y_0+\Delta y)^2\rangle\langle (y_0' + \Delta y')^2\rangle - \langle (y_0+\Delta y)(y_0' + \Delta y')\rangle^2} \\
        &\approx& \sqrt{\left(\langle y_0^2\rangle\langle {y'_0}^2\rangle - \langle y_0 {y'_0}\rangle^2\right) + \left( \langle {y'_0}^2\rangle \langle {\Delta y}^2\rangle+ \langle y_0^2\rangle\langle {\Delta y'}^2\rangle - \langle y'_0 \Delta y \rangle^2 - \langle y_0 {\Delta y'}\rangle^2\right)} \equiv \sqrt{\varepsilon _{y0}^2 + \Delta\varepsilon _{y}^2},
    \end{eqnarray}
\end{widetext}
where we have introduced the difference terms $\Delta y = -2y'_0 L (1+L/l)\delta^2$ and $\Delta y' = -2y'_0 (l/L+1)(1+L/l)\delta^2$. The approximation in the final step assumes that terms of second order in difference terms (e.g., $\Delta y\Delta y'$) are small and can be neglected.
We see that the initial emittance $\varepsilon _{y0}$ is added in quadrature with an emittance growth term $\Delta\varepsilon _{y}$, which can be further simplified to
\begin{widetext}
    \begin{eqnarray}
        \Delta\varepsilon _{y} &=& \sqrt{\langle {y'_0}^2\rangle \langle {\Delta y}^2\rangle+ \langle y_0^2\rangle\langle {\Delta y'}^2\rangle - \langle y'_0 \Delta y \rangle^2 - \langle y_0 {\Delta y'}\rangle^2} \\
        &=& 2\left(1+\frac{L}{l}\right) \sqrt{ L^2 \langle {y'_0}^2\rangle\langle {y'_0}^2 \delta^4\rangle  + \left(\frac{l}{L}+1\right)^2\langle y_0^2\rangle\langle {y'_0}^2 \delta^4\rangle - L^2\langle {y'_0}^2 \delta^2\rangle^2 - \left(\frac{l}{L}+1\right)^2\langle y_0 y'_0 \delta^2\rangle^2} \\
        &=& 2\left(1+\frac{L}{l}\right) \sqrt{ 3 L^2 \sigma_{y'}^4\sigma_{\delta}^4 + 3\left(\frac{l}{L}+1\right)^2\sigma_{y}^2\sigma_{y'}^2\sigma_{\delta}^4 - L^2\sigma_{y'}^4\sigma_{\delta}^4 - 0}  \\
        &=& 2\left(1+\frac{L}{l}\right) \sigma_{\delta}^2 \sqrt{ 2 L^2 \frac{\varepsilon _y^2}{\beta_0^2} + 3\left(\frac{l}{L}+1\right)^2\varepsilon _y^2}
        \hspace{1em} = \hspace{1em} 2\left(1+\frac{L}{l}\right) \sigma_{\delta}^2 \varepsilon _y \sqrt{ 2 \frac{L^2}{\beta_0^2} + 3\left(\frac{l}{L}+1\right)^2}
    \end{eqnarray}
\end{widetext}
where we have defined an initial rms beam size $\sigma_{y} = \langle y_0^2\rangle$ and divergence $\sigma_{y'}=\langle {y'_0}^2\rangle$, used $\langle u^4 \rangle=3\sigma_u^4$ for the $\delta^4$ terms, used that the cross term $\langle y_0 y'_0\rangle$ vanishes  and that $\varepsilon _{y0} = \sigma_{y'} \sigma_{y}$ (true for matched beams where $\alpha_y=0$). This expression can be recast into a relative emittance growth, where we also change from geometric to normalized emittance ($\varepsilon _{n} \approx \gamma\varepsilon )$:
\begin{equation}
    \frac{\Delta\varepsilon _{ny}}{\varepsilon _{ny}} \approx 2\left(1+\frac{L}{l}\right) \sigma_{\delta}^2 \sqrt{ 2 \frac{L^2}{\beta_0^2} + 3\left(\frac{l}{L}+1\right)^2}. 
\end{equation}
In the case where $\beta_0 \ll L$, which is typically the case for the staging lattices considered, we can further simplify to
\begin{equation}
    \frac{\Delta\varepsilon _{nx}}{\varepsilon _{nx}} = \frac{\Delta\varepsilon _{ny}}{\varepsilon _{ny}} \approx \sqrt{8}\left(1+\frac{L}{l}\right) \frac{L}{\beta_0} \sigma_{\delta}^2,
\end{equation}
where we in the first equality have made use of the symmetry between the two transverse planes, i.e.~that the optics functions evolve identically in $x$ and $y$.

\section{Derivation of emittance growth from geometric effects in nonlinear plasma lenses}
\label{app:nonlinear-forces}
To estimate the emittance growth from geometric effects, or nonlinear forces, in the plasma lenses, we will propagate a single particle through a simplified staging lattice. We do not need to consider energy-dependent effects (handled separately in Appendix~\ref{app:second-order-chromaticity}), implying that we can ignore the dipole magnets. This reduces the lattice to three drifts ($L$, $2l$ and $L$) and two plasma lenses, which will both be considered as thin (i.e., a single kick).

Consider a particle with initial positions $x_0$ and $y_0$ in the horizontal and vertical planes, respectively, and initial angles $x'_0$ and $y'_0$. In the plasma lenses, a kick is applied based on the particle position, but at this location the position is dominated by the angles and not the initial position; e.g., in the first lens $x_0+x'_0L \to x'_0L$. The angle term dominates because for a matched beam $\langle x_0 \rangle/\langle x'_0 \rangle = \beta_0$ and in our setup we assume $\beta_0 \ll L$ (and hence $x_0 \ll x'_0L$), which is the reason why chromaticity correction is required in the first place. We can therefore assume that the initial position is negligible.

The particle starts by drifting to the first lens. Here, the offsets will be
\begin{eqnarray}
    x_1 &=& x'_0 L \\
    y_1 &=& y'_0 L,
\end{eqnarray}
where we have dropped the negligible $x_0$ and $y_0$ terms. The lenses have a focal length $f = (L^{-1}+l^{-1})^{-1}$, such that after receiving a nonlinear kick from the lens (see Eqs.~\ref{eq:nonlinear-magnetic-field-x}, ~\ref{eq:nonlinear-magnetic-field-y}, ~\ref{eq:nonlinear-electric-field-x} and \ref{eq:nonlinear-electric-field-y}), the angles will be
\begin{eqnarray}
    x'_1 &=& x'_0-\frac{1}{f}\left(x_1+\tau_x \frac{x_1^2 + y_1^2}{2}\right)\\
    &=& -x'_0\frac{L}{l} -\frac{1}{2} \tau_x L \left(1+\frac{L}{l}\right) \left({x_0'}^2 + {y_0'}^2\right)
\end{eqnarray}
and 
\begin{eqnarray}
    y'_1 &=& y'_0-\frac{1}{f}(y_1+\tau_x x_1 y_1)\\
    &=& -y'_0\frac{L}{l} - \tau_x L \left(1+\frac{L}{l}\right)y'_0 x'_0.
\end{eqnarray}
Next, the particle drifts to the second lens, where the offsets will be
\begin{eqnarray}
    x_2 &=& - L x'_0 - \tau_x L \left({x'_0}^2 + {y'_0}^2\right)(l+L) \\
    y_2 &=& - L y'_0 - 2 \tau_x L y'_0 x'_0 (l+L).
\end{eqnarray}
It then receives a kick from the second lens, after which the angles become
\begin{eqnarray*}
    x'_2 &=& x'_1 -\frac{1}{f}\left(x_2+\tau_x \frac{x_2^2 + y_2^2}{2}\right) \\
    &=& x'_0 + \tau_x L  \left(\frac{l}{L}+1\right)\left({x'_0}^2+{y_0'}^2\right) \\
    && - \tau_x^2  L^2 \left(1+\frac{L}{l}\right)\left(\frac{l}{L}+1\right) x'_0\left({x'_0}^2+3{y'_0}^2\right) 
     + O(\tau_x^3)
\end{eqnarray*}
and
\begin{eqnarray*}
    y'_2 &=& y'_1-\frac{1}{f}(y_2+\tau_x x_2 y_2)\\
    &=& y'_0 + 2 \tau_x L\left(\frac{l}{L}+1\right)  x'_0 y'_0 \\
    && -\tau_x^2 L^2 \left(1+\frac{L}{l}\right) \left(\frac{l}{L}+1\right) y'_0 \left(3{x'_0}^2+{y'_0}^2\right)  
     + O(\tau_x^3),
\end{eqnarray*}
where we have neglected terms of higher than second order in $\tau_x$ [i.e., $\mathcal{O}(\tau_x^3)$].
Finally, the particle drifts to the end of the lattice, resulting in a final offset 
\begin{eqnarray*}
    x_3 &=& - \tau_x^2 L^3 \left(1+\frac{L}{l}\right)\left(\frac{l}{L}+1\right) x'_0 ({x'_0}^2+3{y'_0}^2) + O(\tau_x^3) \\
    y_3 &=& - \tau_x^2 L^3 \left(1+\frac{L}{l}\right) \left(\frac{l}{L}+1\right) y'_0 (3{x'_0}^2+{y'_0}^2) + O(\tau_x^3).
\end{eqnarray*}
The final angles are the same as after the second lens, i.e.~$x'_3 = x'_2$ and $y'_3 = y'_2$. This shows that the $-I$ transform does indeed cancel the geometric effects to first order in the final offset [i.e., there is no $O(\tau_x)$ term], but not on the final angle (which indeed has a $O(\tau_x)$ term).

Starting with the assumption that the initial distributions in all phase-space variables ($x_0$, $y_0$, $x'_0$, $y'_0$) are Gaussian, we can estimate the rms geometric emittance (first in the horizontal plane only) to be
\begin{eqnarray*}
    \varepsilon _{x} &=& \sqrt{\langle x_3^2\rangle\langle {x'_3}^2\rangle - \langle x_3 {x'_3}\rangle^2} \\
    &\approx& \sqrt{\langle (x_0+\Delta x)^2\rangle\langle {x'_0}^2\rangle - \langle (x_0+\Delta x) x'_0\rangle^2} \\
    &\approx& \sqrt{\left(\langle x_0^2\rangle\langle {x'_0}^2\rangle - \langle x_0 {x'_0}\rangle^2\right) + \left(\langle \Delta x^2\rangle\langle {x'_0}^2\rangle - \langle \Delta x {x'_0}\rangle^2\right)} \\
    &\approx& \sqrt{\varepsilon _{x0}^2 + \left(\langle \Delta x^2\rangle\langle {x'_0}^2\rangle - \langle \Delta x {x'_0}\rangle^2\right)} \equiv \sqrt{\varepsilon _{x0}^2 +\Delta\varepsilon _x^2},
\end{eqnarray*}
where in the second step, we make use of the finding that the increase in phase-space area is dominated by the positional offsets and not the angular offsets. Next, we substitute the expression for $\Delta x$ (see $x_3$ above), and define the divergence $\sigma_{x'} = \sqrt{\langle {x'_0}^2\rangle}$, resulting in
\begin{widetext}
    \begin{eqnarray}
        \Delta\varepsilon _x &\approx& \sqrt{\langle \Delta x^2\rangle\langle {x'_0}^2\rangle - \langle \Delta x {x'_0}\rangle^2} \\
        &\approx& \tau_x^2 L^3 \left(1+\frac{L}{l}\right)\left(\frac{l}{L}+1\right) \sqrt{\left\langle {x'_0}^2 \left({x'_0}^2+3{y'_0}^2\right)^2\right\rangle\langle {x'_0}^2\rangle - \left\langle ({x'_0}^2+3{y'_0}^2)  {x'_0}^2\right\rangle^2} \\
        &\approx& \tau_x^2 L^3 \left(1+\frac{L}{l}\right)\left(\frac{l}{L}+1\right)\sigma_{x'} \sqrt{\langle {x'_0}^6\rangle + 6\langle {x'_0}^4{y'_0}^2\rangle + 9\langle {x'_0}^2{y'_0}^4\rangle - \left(\frac{\langle {x'_0}^4\rangle+3\langle {y'_0}^2{x'_0}^2\rangle}{\sigma_{x'}}\right)^2}\\
        &\approx& \tau_x^2 L^3 \left(1+\frac{L}{l}\right)\left(\frac{l}{L}+1\right)\sigma_{x'} \sqrt{15\sigma_{x'}^6 + 18\sigma_{x'}^4\sigma_{y'}^2 + 27\sigma_{x'}^2\sigma_{y'}^4 - 9\sigma_{x'}^6-18\sigma_{x'}^4\sigma_{y'}^2-9\sigma_{x'}^2\sigma_{y'}^4} \\
        &\approx& \tau_x^2 L^3 \left(1+\frac{L}{l}\right)\left(\frac{l}{L}+1\right)\sigma_{x'}^2 \sqrt{6\sigma_{x'}^4 + 18\sigma_{y'}^4} \\
        &\approx& \tau_x^2 L^3 \left(1+\frac{L}{l}\right)\left(\frac{l}{L}+1\right)\frac{\varepsilon _x}{\beta_0^2} \sqrt{6\varepsilon _x^2 + 18\varepsilon _y^2},
    \end{eqnarray}
\end{widetext}
where we used $\langle u^4 \rangle = 3\sigma_u^4$ and $\langle u^6 \rangle = 15\sigma_u^6$ (fourth step) and $\sigma_{x'} = \varepsilon _x/\beta_0$ (last step). Converting to normalized emittance $\varepsilon _{nx} \approx \gamma\varepsilon _{x}$, we get
\begin{equation}
    \frac{\Delta\varepsilon _{nx}}{\varepsilon _{nx}} \approx \frac{\tau_x^2 L^3}{\beta_0^2 \gamma} \left(1+\frac{L}{l}\right)\left(\frac{l}{L}+1\right)\sqrt{6\varepsilon _{nx}^2 + 18\varepsilon _{ny}^2}.
\end{equation}

For the other transverse plane ($y$), we notice that the expressions for $y_3$ and $x_3$ above are symmetric in $x'_0$ and $y'_0$, which means that the derivation for the emittance growth will be identical (but swapping $x$ and $y$). By symmetry, therefore 
\begin{equation}
    \frac{\Delta\varepsilon _{ny}}{\varepsilon _{ny}} \approx \frac{\tau_x^2 L^3}{\beta_0^2 \gamma} \left(1+\frac{L}{l}\right)\left(\frac{l}{L}+1\right)\sqrt{18\varepsilon _{nx}^2 + 6\varepsilon _{ny}^2}
\end{equation}
is the normalized emittance growth in the vertical plane.

\section{Derivation of the induced action from plasma-lens offsets}
\label{app:induced-action}
Following a similar method as in Appendix~\ref{app:second-order-chromaticity} and \ref{app:nonlinear-forces} above, the normalized phase-space offset or \textit{action} (defined in Eq.~\ref{eq:action}) can be calculated. Here, we assume that the nonlinear (i.e., higher-order) plasma-lens fields and the central sextupole are all negligible compared to the linear (i.e., first-order) plasma-lens fields.

Restricting ourselves to the horizontal plane (without loss of generality), we start by defining the offsets $\Delta_1$ and $\Delta_2$ of the first and second lens, respectively, and aim to find the resulting final offset $\Delta x$ and angle $\Delta x'$ of the beam. The beam starts with zero offset and angle: 
\begin{equation}
    x_0 = x'_0 = 0.
\end{equation}
Therefore, after a drift of length $L$ (just before the first lens), we still have zero offset
\begin{equation}
    x_1 = 0.
\end{equation}
However, after the first lens, which has focal length $f = (L^{-1}+l^{-1})^{-1}$, an angle is introduced:
\begin{equation}
    x'_1 = \frac{\Delta_1}{f} = \Delta_1\left(\frac{1}{L}+\frac{1}{l}\right),
\end{equation}
where $l$ is the distance from the lens to the lattice center. Note that positive lens offsets produce positive angles, whereas positive particle offsets (for a zero lens offset) produce negative angles. Transporting this to the second lens, the offset becomes
\begin{equation}
    x_2 = x_1 + 2l x'_1 = 2\Delta_1\left(\frac{l}{L}+1\right).
\end{equation}
The resulting angle after the second lens, which is also the final angle, is then
\begin{eqnarray*}
    \Delta x' &=& x'_2 = x'_1 + (\Delta_2-x_2)/f \\
    &=& \Delta_1\left(\frac{1}{L}+\frac{1}{l}\right) + \left(\Delta_2-2\Delta_1\left(\frac{l}{L}+1\right)\right)\left(\frac{1}{L}+\frac{1}{l}\right) \\
    &=& \left(\Delta_2-\Delta_1\left(\frac{2l}{L}-1\right)\right)\left(\frac{1}{L}+\frac{1}{l}\right).
\end{eqnarray*}
Lastly, the final offset becomes
\begin{eqnarray*}
    \Delta x &=& x_2 + Lx'_2 \\
    &=& 2\Delta_1\left(\frac{l}{L}+1\right) + \left(\Delta_2-\Delta_1\left(\frac{2l}{L}-1\right)\right)\left(1+\frac{L}{l}\right) \\
    &=& 2\Delta_1\frac{l}{L}\left(1+\frac{L}{l}\right) + \left(\Delta_2-\Delta_1\left(\frac{2l}{L}-1\right)\right)\left(1+\frac{L}{l}\right) \\
    &=& \left(\Delta_1+\Delta_2\right)L\left(\frac{1}{L}+\frac{1}{l}\right). \\
\end{eqnarray*}

As a brief aside, since the two lens offsets affect the final offset and angle differently, this means that lens offsets can be used to correct the orbit into the subsequent stage. We can represent the effect in matrix form:
\begin{equation}
    \begin{bmatrix} \Delta x\\ \Delta x' \end{bmatrix} = \left(\frac{1}{L}+\frac{1}{l}\right)
    \begin{bmatrix} L & L \\ \left(1-\frac{2l}{L}\right) & 1 \end{bmatrix} 
    \begin{bmatrix} \Delta_1 \\ \Delta_2 \end{bmatrix}.
\end{equation}
This matrix can be inverted to determine how to move the lenses in tandem to correct a given offset in phase space:
\begin{equation}
    \begin{bmatrix} \Delta_1 \\ \Delta_2 \end{bmatrix} 
    = \frac{-1}{2(L+l)}
    \begin{bmatrix} L & -L^2 \\ 2l-L & L^2 \end{bmatrix} 
    \begin{bmatrix} \Delta x\\ \Delta x' \end{bmatrix},
\end{equation}
where a negative sign has been introduced such that the phase-space offset we wish to mitigate is subtracted instead of added. The same matrix will also apply in the $y$ direction.

At last, we can calculate the induced action, by substituting in $\Delta x$ and $\Delta x'$ to get
\begin{widetext}
\begin{eqnarray}
    J_x &=& \frac{\gamma}{2}\left(\frac{1}{\beta_0}\Delta x^2 + \beta_0 {\Delta x'}^2 \right) = \frac{\gamma}{2}\left(\frac{1}{\beta_0}\left(\left(\Delta_1+\Delta_2\right)L\right)^2 + \beta_0 \left( \Delta_2-\Delta_1\left(\frac{2l}{L}-1\right)  \right)^2  \right)\left(\frac{1}{L}+\frac{1}{l}\right)^2 \\
    &=& \frac{\gamma\beta_0}{2}\left(\left(\frac{L}{\beta_0}\right)^2(\Delta_1^2+2\Delta_1\Delta_2+\Delta_2^2)+\Delta_2^2-2\Delta_1\Delta_2\left(\frac{2l}{L}-1\right)+\Delta_1^2\left(\frac{2l}{L}-1\right)^2\right)\left(\frac{1}{L}+\frac{1}{l}\right)^2 \\
    &\approx& \frac{\gamma}{2\beta_0}(\Delta_1^2+2\Delta_1\Delta_2+\Delta_2^2)\left(1+\frac{L}{l}\right)^2,
\end{eqnarray}
\end{widetext}
where the final approximation assumes that $\beta_0 \ll L$. Further, we can assume that the offset of the first and second lens are random, normally distributed with the same rms $\sigma_{\Delta x} = \sqrt{\langle \Delta_1\rangle} = \sqrt{\langle \Delta_2\rangle}$, and not correlated with each other (i.e., cross terms $\Delta_1\Delta_2$ vanish). The mean value of the induced action then becomes
\begin{equation}
    \langle J_x \rangle \approx \frac{\gamma}{\beta_0}\sigma_{\Delta x}^2\left(1+\frac{L}{l}\right)^2. \\
\end{equation}
Converting this into an estimated misalignment tolerance, where we limit the induced action to be less than the emittance, i.e.~$\langle J_x \rangle < \varepsilon _{nx}$, we solve for $\sigma_{\Delta x}$ to get
\begin{equation}
    \sigma_{\Delta x}^{\max} < \sqrt{\frac{\varepsilon _{nx}\beta_0}{\gamma}}\left(1+\frac{L}{l}\right)^{-1} = \sigma_{x0}\left(1+\frac{L}{l}\right)^{-1} ,
\end{equation}
where we have substituted in the initial beam size $\sigma_{x0}$. In summary, the plasma-lens offset tolerance is simply a fraction of the beam size; about a two thirds (or $0.65\sigma_{x0}$ exactly) for the example given in Table~\ref{tab:working-point}.


\end{document}